\documentclass[pre,aps,twocolumn,superscriptaddress,amsmath,showpacs,floatfix]{revtex4}

\usepackage{epsfig}
\usepackage{amssymb}

\begin{document}
\title{Statistical mechanics of bilayer membrane with a fixed
  projected area}
\author{O. Farago}
\email{farago@mrl.ucsb.edu}
\affiliation{
  Materials Research Laboratory, University of
  California, Santa Barbara, CA 93106}
\affiliation{
  Department of Physics, Korea Advanced Institute of
  Science and Technology (KAIST), 373-1 Kusong-dong,
  Yusong-gu, Taejon 305-701, South Korea.}
\author{P. Pincus}
\affiliation{
  Materials Research Laboratory, University of
  California, Santa Barbara, CA 93106}
\affiliation{
  Department of Physics, Korea Advanced Institute of
  Science and Technology (KAIST), 373-1 Kusong-dong,
  Yusong-gu, Taejon 305-701, South Korea.}
\date{\today}

\begin{abstract}
  \vspace{0.5cm}
  The equilibrium and fluctuation methods for determining the surface
  tension, $\sigma$, and bending modulus, $\kappa$, of a bilayer
  membrane with a fixed projected area are discussed. In the
  fluctuation method the elastic coefficients $\sigma$ and $\kappa$
  are measured from the amplitude of thermal fluctuations of the
  planar membrane, while in the equilibrium method the free energy
  required to deform the membrane is considered. The latter approach
  is used to derive new expressions for $\sigma$ and $\kappa$ (as well
  as for the saddle-splay modulus), which relate them to the
  pair-interactions between the amphiphiles forming the membrane. We
  use linear response theory to argue that the two routes lead to
  similar values for $\sigma$ and $\kappa$. This argument is confirmed
  by Monte Carlo simulations of a model membrane whose elastic
  coefficients are calculated using both methods.
\end{abstract}
\maketitle

\begin{widetext}
\section{Introduction}
\label{intro}

The Bilayer membrane, a double sheet of surfactants separating two
aqueous phases, is one of the structures formed by the self-assembly
of amphiphilic molecules in water \cite{israelachvili}. The driving
force in this process is the hydrophobic effect which favors exposing
the hydrophilic part of the molecules to the water while shielding the
``oily'' part from aqueous contact \cite{tanford,gelbart}. The ongoing
interest in such membranes is due to many reasons, among which are
their predominant role in the organization of the biological cells
\cite{alberts}, and their various applications in many industrial
sectors \cite{rosen}. Bilayer amphiphilic sheets have very special
mechanical properties: While being strongly resistant to lateral
mechanical stretching or compression, they are highly flexible and can
exhibit large thermally excited undulations
\cite{safran,lipowsky_sackmann}. This unique elastic behavior, namely
the stability against external perturbations on the one hand, but the
ease in going from one shape to another on the other hand, is
important for the activity of living cells \cite{gennis}.
Consequently, there has been a great effort to understand the
elasticity of bilayer systems
\cite{safran,lipowsky_sackmann,winter_school,petrov}.

Bilayer membranes are quasi two-dimensional (2D) objects: their
thickness is typically of the size of a few nanometers (roughly, twice
the length of the constituent amphiphilic molecules), while their
lateral extension can reach up to several micrometers. Since the
membrane appears as a thin film on the mesoscopic scale, its physical
properties are often studied using coarse-grained phenomenological
models treating the membrane as a smooth continuous 2D sheet
\cite{safran,lipowsky_sackmann,winter_school,gompper_schick}. Membrane
elasticity has been traditionally studied using the Helfrich effective
surface Hamiltonian which relates the elastic energy to the local
principle curvatures of the membrane $c_1$ and $c_2$, and which has
the following form \cite{helfrich}:
\begin{equation}
{\cal H}=\int_{A} dS\,
\left[\sigma_0+\frac{1}{2}\kappa_0
\left(J-2c_{0}\right)^2+{\bar{\kappa}}_0K\right],
\label{helfhamiltonian}
\end{equation}
where $J\equiv c_1+c_2$ and $K\equiv c_1c_2$ are the total and
Gaussian curvatures respectively. The integration in
Eq.(\ref{helfhamiltonian}) is carried over the whole surface of the
membrane. The Helfrich Hamiltonian is derived by assuming that local
curvatures are small, and the free energy can be expanded to second
order in $J$ and to first order in $K$. It, therefore, involves four
phenomenological parameters: the spontaneous curvature $c_0$, and three
elastic coefficients - the surface tension $\sigma_0$, the bending
modulus $\kappa_0$, and the saddle-splay modulus $\bar{\kappa}_0$,
whose values depend on the area density of the amphiphiles. If the
number of these is fixed, then one should also consider the
corrections to Hamiltonian (\ref{helfhamiltonian}) due to the changes
in the area of the fluctuating membrane. For weakly fluctuating
membranes these corrections can be assumed to be small. The surface
tension, which is usually associated with the free energy cost for
adding molecules to the membrane (at a fixed density), is related in
the case of membranes with fixed number of amphiphiles to the
area-density dependent (Schulman) elastic energy
\cite{schulman,degennes_taupin,farago3}.

The Helfrich Hamiltonian has been very successful in describing the
shape and the phase diagram of complex interfaces
\cite{miao,seifert1,wortis}. It also yields a correct description of
the thermal fluctuations around the equilibrium surface state
\cite{helfrich_servuss,scott_safran,seifert2}, and of the entropic
forces between membranes \cite{helfrichint}. Because it is
phenomenological, the Helfrich Hamiltonian provides no information
about the values of the elastic coefficients.  Many theories have been
developed that attempt to relate the elastic coefficients introduced
by the Helfrich Hamiltonian to microscopic entities and the
interactions between them
\cite{petrov_bivas,szleifer,andelman,chacon,magnus}.  In fact, these
theories are usually concerned with the free energy of the surface,
rather than the Hamiltonian. The free energy is assumed to have the
same form as the Helfrich Hamiltonian and, hence, usually called the
{\em Helfrich free energy}\/ (see a more detailed discussion in
section \ref{routes}). The coefficients appearing in the expression
for the free energy, which we denote by $\sigma$, $\kappa$, and
$\bar{\kappa}$, are also referred to as the surface tension, the
bending modulus, and the saddle-splay modulus, respectively. Despite
the similarity in names, there is a significant difference between the
Hamiltonian coefficients (with the subscript 0) and the free energy
coefficients. The former are ``material properties'' which depend on
the internal (potential) energy of the surface. The latter, on the
other hand, are thermodynamic quantities and, as such, are also
influenced by the entropy associated with the thermal fluctuations of
the system.  Their values, therefore, may also depend on the
temperature and the size of the system.

In addition to the above mentioned theories, there has been also an
effort to analyze the elastic behavior in the context of the
thermodynamics and statistical mechanics of curved interfaces
\cite{rowlinson_widom,schofield_henderson,kozlov_markin,blokhuis_bedeaux1,robledo_varea,oversteegen}.
The last approach has the potential of providing exact ``virial''
expressions for $\sigma$, $\kappa$, and $\bar{\kappa}$ in terms of the
microscopic forces between the amphiphiles and the pair distribution
function. One of the systems whose statistical mechanics has been
studied extensively is that of a simple liquid-vapor interface.
Although this seems to be a rather simple system, the determination of
its elastic moduli is quite complicated and involves a set of
technical and conceptual problems. Below we discuss some of them:
\begin{itemize}
\item{One problem is related to the finite thickness of the interface,
    namely to the fact that the local concentration is not a step
    function but changes gradually while going from one phase to the
    other. Consequently, there is some ambiguity about the location of
    the dividing plane that separates the two phases and to which the
    Helfrich Hamiltonian is applied. It turns out that the values of
    the rigidity constants $\kappa$ and $\bar{\kappa}$ (the
    coefficients of the second order terms in the curvatures $c_1$ and
    $c_2$) depend on the choice of the dividing surface
    \cite{rowlinson}. The dependence of the rigidity constants on the
    reference surface had led people to question the validity of
    continuing the Helfrich free energy expansion beyond the linear
    term in curvature. This problem has been recently tackled by van
    Giessen and Blokhuis \cite{giessen_blokhuis} who used computer
    simulation to determine the rigidity constants of a curved
    liquid-vapor interface in a system of particles interacting via a
    truncated Lennard-Jones (LJ) potential. They have demonstrated
    that although one needs to state which convention for locating the
    dividing surface is used when providing the values of $\kappa$ and
    $\bar{\kappa}$, this fact does not render the Helfrich free energy
    useless, nor does it diminish the importance of these quantities
    in describing the elastic properties of the interface.}
\item{A second problem that makes the determination of the rigidity
    constants difficult is a technical one: In their paper van Giessen
    and Blokhuis used the virial expressions given in
    Ref.~\cite{blokhuis_bedeaux2} to evaluate the values of $\kappa$
    and $\bar{\kappa}$. These expressions relate the rigidity
    constants to the derivative of the pair density distribution
    function with respect to the radius of curvature $R_c$. This means
    that the values of the rigidity constants of a {\em planar}
    interface cannot be determined from the simulation of that system
    only, but it is necessary to perform a set of simulations of
    curved interfaces with very large values of $R_c$. For the
    interfaces investigated in Ref.~\cite{giessen_blokhuis}, it turns
    out that in the large $R_c$ regime the dependence of the pair
    density function on $R_c$ is very weak. Consequently, it was
    impossible to determine $\kappa$ and $\bar{\kappa}$ accurately,
    and only a rough estimate of these quantities could be obtained.}
\item{A third problem, a more fundamental one, is related to the
    method of calculating the rigidity constants $\kappa$ and
    $\bar{\kappa}$ and to our interpretation of their physical
    meaning. The theoretical and experimental methods for determining
    the elastic coefficients of interfaces can be classified into {\em
      equilibrium}\/ (or {\em mechanical}\/) methods and {\em
      fluctuation} methods \cite{blokhuis_bedeaux3,seifert_lipowsky}.
    The difference between these two approaches is in the context in
    which the Helfrich Hamiltonian and the associated free energy are
    used: In the equilibrium approach one extracts the elastic
    coefficients by comparing the free energies of two equilibrium
    surfaces with different curvatures. In the fluctuation approach,
    on the other hand, the Helfrich Hamiltonian is used to calculate
    the free energy cost due to a thermal fluctuation that changes the
    local curvature from its equilibrium value. The elastic
    coefficients are derived from the mean-square amplitudes of the
    fluctuations. The situation in which there exist two methods for
    calculating elastic moduli is reminiscent of other cases, for
    instance, the two different methods of evaluating the elastic
    constants of thermodynamic systems in linear elasticity theory
    \cite{shh,parrinello_rahman,farago_kantor,farago}, and the two
    approaches for determining the surface tension of a planar
    interface \cite{kirkwood_buff,triezenberg_zwanzig}. In the latter
    examples the different approaches lead to the same values for the
    mechanical moduli, in accord with the {\em linear response
      theory}\/\cite{plischke,chaikin_lubensky}. This is not the case
    with the rigidity constants of a liquid-vapor interface
    \cite{blokhuis_bedeaux3}. The discrepancy between the two methods
    is due to the fact that in order to change the equilibrium radius
    of curvature of, say, a spherical liquid drop, it is necessary to
    change its volume as well. This means a change in the volume
    fractions of the two phases (i.e., the condensation of vapor or
    the evaporation of liquid), and it thus requires the variation of
    the thermodynamic variables like the temperature or the chemical
    potential. In the fluctuation case the radius of curvature is
    varied by thermal fluctuations, while the thermodynamic variables
    are not altered.}
\end{itemize}

In this paper we discuss the statistical mechanics of fluid bilayer
membranes. We derive expressions for the elastic coefficients
$\sigma$, $\kappa$, and $\bar{\kappa}$ of the membranes, relating them
to the interactions and the correlation functions between the
amphiphiles forming the bilayer. We use these expressions for a Monte
Carlo (MC) determination of the elastic coefficients of a bilayer
membrane computer model. Unlike the expressions derived for the
rigidity constants of a liquid-vapor interface, our expressions are
such that they can be evaluated using a {\em single}\/ MC run
performed on the (quasi) flat membrane reference system only. This
feature greatly simplifies the computational procedure, and makes it
more efficient and well-controlled. Another important distinction
between the membranes discussed in this paper and the system of
liquid-vapor interface studied in Ref.~\cite{giessen_blokhuis} is the
fact that the mechanical and the fluctuation methods for determining
their rigidity constants lead to similar results. Our expressions are
derived using the mechanical approach, namely by calculating the free
energy variations resulting from the change in the area and curvature
of the membrane. The numerical values of the elastic coefficients
which we obtain from these expressions are compared with the values
extracted from a spectral analysis of the thermal fluctuations around
the flat reference state.  We find a very good agreement between the
two methods.  This agreement, which is expected by virtue of linear
response theory (see discussion in section \ref{routes}), reflects the
fact that the curvature of the membrane can be varied by changing the
shape of the container (namely, by the application of external forces)
without affecting the thermodynamic properties of the bulk aqueous
phases surrounding it. It should be noted that the experimental values
of $\kappa$ measured (for the same lipid bilayers) using mechanical
and fluctuation methods can differ by as much as a factor of 3
\cite{seifert_lipowsky}. The origin of these discrepancies is not well
understood.

The bilayer computer model which we use in this paper has been
recently introduced by one of us \cite{farago2}. (Here we use a
slightly modified version of that model which we describe in section
\ref{numerical}.) This model has two features which simplify the
derivation of thermodynamic expressions for the elastic coefficients
and the simulations performed for the calculation of these
expressions. First, the simulations are conducted with no solvent
present in the simulations cell, i.e., as if the membrane is in
vacuum. This feature greatly reduces the number of atoms in the
simulation cell, thus enabling us to simulate a relatively large
membrane over a very long MC run. The ability to perform long MC runs
is very important since the quantities whose thermal averages we try
to evaluate are very ``noisy'', and accurate results can be obtained
only if they are measured for a large number of configurations. The
other feature is the nature of the interactions between the molecules
forming the membrane.  In our computer model the amphiphilic molecules
are modeled as trimers and the interactions between their constituent
atoms are pair-wise additive. For such systems the derivation of
expression for the elastic coefficients (see section
\ref{expressions}) is simpler than for systems including many-body
potentials. Our discussion in this paper is, therefore, restricted to
central force systems only.

The paper is organized in the following way: The theoretical aspects
of our study are presented in sections \ref{routes} and
\ref{expressions}. In section \ref{routes} we describe the relation
between the equilibrium and the fluctuation routes for determining the
surface tension $\sigma$ and the bending modulus $\kappa$ of bilayer
membranes, and explain why these methods (if used appropriately) lead
to similar results. Then, in section \ref{expressions}, we derive
expressions for these quantities based on the equilibrium approach.
Our expressions relate $\sigma$ and $\kappa$ to the interactions and
the correlation functions between the ``interaction sites'' of the
amphiphilic molecules. The numerical results are presented in section
\ref{numerical} where we calculate the elastic coefficients of our
model system using the two methods and find a very good agreement
between them. Some technical aspects of the simulations are discussed
in the Appendix.  Finally we conclude in section \ref{summary}.

\section{The equilibrium and fluctuation routes to membrane elasticity} 
\label{routes}

Linear response is a fundamental theorem which relates the
fluctuations of a system around its equilibrium state and the response
of the system to weak perturbations \cite{plischke,chaikin_lubensky}.
In the context of elasticity theory it provides a link between the
shape fluctuations of thermodynamic systems and their elastic moduli.
For example, when a 2D {\em flat}\/ interface is slightly stretched or
compressed from its equilibrium area $A_0$, the variation of the
(small) surface pressure $\Pi$ is given by
\cite{landau_lifshits,wallace}
\begin{equation}
K_A=-A_0\frac{\partial \Pi}{\partial A},
\end{equation}
where $A$ is the area of the interface and $K_A$ is the
stretching/compression modulus. The above relation provides one way to
measure $K_A$. An alternative approach for measuring $K_A$ is to
consider the thermal fluctuations of the area $A$ around the
equilibrium area $A_0$ \cite{parrinello_rahman,lindahl_edholm}.  The
equipartition theorem suggests that in the low temperature limit when
fluctuations around $A_0$ are small
\begin{equation}
\langle (A-A_0)^2\rangle=\frac{k_BTA}{K_A},
\end{equation}
where $k_B$ is the Boltzmann constant and $T$ is the temperature,
while $\langle\cdots \rangle$ denotes a thermal average. Linear
response theory can be also applied to describe the normal,
curvature-forming, fluctuations of the 2D interface. The discussion in
this case (of normal fluctuations) is, however, somewhat more
complicated. A proof of the equivalence between the equilibrium and
the fluctuation routes to the {\em surface tension}\/ $\sigma$ of a
fluctuating interface had been presented with great clarity by Cai et
al.~\cite{cai}. Below we extend that proof and address the two routes
to the {\em bending modulus}\/ $\kappa$ as well. One important
difference between the present discussion and the one presented in
Ref.~\cite{cai} is related to the nature of the fluctuating surfaces
in question. Here, we consider an elastic surface consisting of a
fixed number of molecules whose area density is varied when it
fluctuates. By contrast, the surface studied in Ref.~\cite{cai} is
incompressible and its area density is fixed to its equilibrium value.
The variation of the total area of the latter is achieved via the
exchange of molecules between the surface and the embedding solvent. A
more detailed discussion on the differences between the elastic
properties of compressible and incompressible surfaces appears in
Ref.~\cite{farago3}.

Let us consider a 2D surface that spans a planar frame of a total area
$A_p$ which does {\em not}\/ necessarily coincide with the equilibrium
(Schulman) area $A_0$. The surface is free to undulate in the
direction normal to frame. The ensemble of conformations which the
surface attains is governed by a Hamiltonian ${\cal
  H}\left(h(\vec{r})\right)$ relating the elastic energy to the
conformation of the surface. The conformation of the surface is
described by some ``gauge'' function $h(\vec{r})$, where
$\vec{r}=(x,y)$ label the points on the reference surface. The exact
form of the Hamiltonian ${\cal H}$ is unimportant and, in particular,
it is not limited to the Helfrich form (\ref{helfhamiltonian}). As we
are interested in moderately-fluctuating surfaces (with no overhangs),
we shall use the the so called Monge gauge $z=h(\vec{r})$, where $h$
is the height of the surface above the frame reference plane. In what
follows we will restrict our discussion to symmetric surfaces (such as
bilayers) with no spontaneous curvature, i.e., with no preference to
bend toward either the ``upper'' or ``lower'' side of the surface. In
other words, we assume that the average conformation of the surface is
flat and for each $\vec{r}$
\begin{equation}
\langle h\left(\vec{r}\right)\rangle=0.
\label{nospontan}
\end{equation}
We also assume that the surface under consideration is mechanically
stable, and that the validity of Eq.(\ref{nospontan}) is not due to
the partition of the configurations phase space into several 
sub spaces for which $\langle h\left(\vec{r}\right)\rangle\neq 0$.

If the frame (projected) area $A_p$ is not equal to the equilibrium
area $A_0$ then it is necessary to apply a tangential surface pressure
in order to fix the area of the frame. If, in addition, normal forces
are applied then relation (\ref{nospontan}) breaks down. The function
\begin{equation}
\bar{h}(\vec{r})=\langle h\left(\vec{r}\right)\rangle
\label{hbar}
\end{equation}
can be regarded as the strain field describing the deformed state of
the surface. The free energy of a system subjected to a small
deformation can be expanded in a power series in the strain variables.
In full analogy to Hamiltonian (\ref{helfhamiltonian}), we can write
the Helfrich free energy of the surface in the following form:
\begin{equation}
F\left(\bar{h}\right)=F\left(\bar{h}=0\right)
+\sigma \left(A\left(\bar{h}\right)-A_p\right) 
+ \frac{1}{2}\kappa 
{\bar{J}}^{\,2}\left(\bar{h}\right) 
+ \bar{\kappa} \bar{K}\left(\bar{h}\right)
+ {\rm h.o.t},
\label{helffenergy}
\end{equation}
where $A\left(\bar{h}\right)$ is total area of the surface defined by
the function $\bar{h}(\vec{r})$, while $\bar{J}\left(\bar{h}\right)$
and $\bar{K}\left(\bar{h}\right)$ denote, respectively, the {\em
  integrated}\/ total and Gaussian curvatures defined by
\begin{equation}
{\bar{J}}^{\,2}\equiv \int_{A_p} d\vec{r}\, 
J^{\,2}\left(\bar{h}(\vec{r})\right),
\label{intcurv}
\end{equation}
and 
\begin{equation}
\bar{K}\equiv \int_{A_p} d\vec{r}\, 
K\left(\bar{h}(\vec{r})\right).
\label{defgauss}
\end{equation}
In Eq.(\ref{helffenergy}) we set the spontaneous curvature $c_0=0$
[see Eqs.(\ref{nospontan}) and (\ref{hbar})], and use the {\em
  effective}\/ ({\em normalized}) values of the elastic coefficients
which are different from the ``bare'' values appearing in the
Hamiltonian (\ref{helfhamiltonian}) (see discussion earlier in section
\ref{intro}). The higher order terms (h.o.t) in Eq.(\ref{helffenergy})
include both products of the small variables $(A-A_p)/A_p$,
$\bar{J}^{\, 2}$, and $\bar{K}$, as well as terms involving the {\em
  gradients}\/ of the local curvatures. The latter are assumed to be
small since we consider only nearly-flat surfaces described by
functions $\bar{h}$ which vary slowly in space.  Since $\sigma$,
$\kappa$, and $\bar{\kappa}$ appear as the coefficients of the free
energy expansion in strain variables [Eq.(\ref{helffenergy})], they
can be also related to the following partial derivatives
\begin{equation}
\sigma =\frac{\partial F}{\partial A}
\Biggm|_{\bar{h}(\vec{r})=0},
\label{mechstension}
\end{equation}
\begin{equation}
\kappa =\frac{\partial^2 F}{\partial {\bar{J}}^{\,2}}
\Biggm|_{\bar{h}(\vec{r})=0},
\label{mechbendmod}
\end{equation}
and
\begin{equation}
\bar{\kappa} =\frac{\partial F}{\partial \bar{K}}
\Biggm|_{\bar{h}(\vec{r})=0},
\label{mechgauss}
\end{equation}
evaluated at the reference state $\bar{h}(\vec{r})=0$.

Equations (\ref{mechstension})--(\ref{mechgauss}) express the
equilibrium (mechanical) route to $\sigma$, $\kappa$, and
$\bar{\kappa}$. The complementary fluctuations approach is more easily
formulated in Fourier rather than in real space. Let us take a square
frame (the reference surface) of linear size $L_p=\sqrt{A_p}$, and
discretized it into $N^2$ square cells (``patches'') of linear size
$l=L_p/N$, where $l$ is some microscopic length of the order of the
size of the constituent molecules. Since the description of the
membrane as a 2D continuous sheet breaks down on length scales below
$l$, the surface has to be defined only over a discrete set of points
$\left\{\vec{r}_g=\left(x_g,y_g\right)\right\}$ each of which located
in the center of a grid cell. Outside the frame region, the function
can be defined by periodic extension of period $L_p$, i.e.
$\bar{h}\left(x_g+n_1L_p,y_g+n_2L_p\right)=\bar{h}\left(x_g,y_g\right)$
where $n_1$ and $n_2$ are integer numbers. The Fourier transform of
the (real) function $\bar{h}(\vec{r}_g)$ is defined by
\begin{equation}
\bar{h}_{\vec{q}}=\frac{l}{L_p}\sum_{\vec{r}_g}
\bar{h}\left(\vec{r}_g\right)e^{-i\vec{q}\cdot\vec{r}_g}.
\label{invtransform}
\end{equation}
where the two dimensional wave-vector $\vec{q}$ has $N^2$ discrete
values satisfying
\begin{equation}
\left\{q_x,q_y=2\pi m/L_p,\ 
  m=-N/2,\ldots,N/2-1\right\}.
\end{equation}
The inverse transform is given by
\begin{equation}
\bar{h}(\vec{r}_g)=\frac{l}{L_p}\sum_{\vec{q}}
\bar{h}_{\vec{q}}\, e^{i\vec{q}\cdot\vec{r}_g},
\label{transform}
\end{equation}
If the topology of the surface is fixed and it does not form
``handles'' then the periodicity of the surface leads to the vanishing
of the Gaussian curvature (\ref{defgauss}) (Gauss-Bonnet theorem).
Writing the expressions for the area $A\left(\bar{h}\right)$ and the
integrated total curvature $\bar{J}$ in terms of Fourier coordinates:
\begin{equation}
A\left(\bar{h}\right)=A_p+\frac{l^2}{2}
\sum_{\vec{q}}q^2\bar{h}_{\vec{q}}\bar{h}_{-\vec{q}}+{\cal O}
\left(|\bar{h}_{\vec{q}}|^4\right)
\end{equation}
and
\begin{equation}
\bar{J}^{\, 2}\left(\bar{h}\right)=l^2
\sum_{\vec{q}}q^4\bar{h}_{\vec{q}}\bar{h}_{-\vec{q}}+{\cal O}
\left(|\bar{h}_{\vec{q}}|^4\right),
\end{equation}
and substituting them in Eq.(\ref{helffenergy}), we obtain the
following expression for the free energy
\begin{equation}
F\left(\bar{h}\right)=F\left(\bar{h}=0\right)+\frac{l^2}{2}
 \sum_{\vec{q}}\left[\sigma q^2+\kappa q^4 + {\cal O}(q^6) 
\right]\bar{h}_{\vec{q}}\bar{h}_{-\vec{q}}
+{\cal O} \left(|\bar{h}_{\vec{q}}|^4\right).
\label{fourierfenergy}
\end{equation}

The free energy (\ref{fourierfenergy}) can be related to the surface
Hamiltonian ${\cal H}\left(\left\{h(\vec{r}_g)\right\}\right)$ via the
partition function. We may use the Fourier transform 
\begin{equation}
h_{\vec{q}}=\frac{l}{L_p}\sum_{\vec{r}_g}
h\left(\vec{r}_g\right)e^{-i\vec{q}\cdot\vec{r}_g}
\label{invtransform2}
\end{equation}
of the function $h\left(\left\{\vec{r}_g\right\}\right)$, and express
the Hamiltonian as a function of the Fourier modes: ${\cal
  H}\left(\left\{h_{\vec{q}}\right\}\right)$. Introducing the set of
Lagrange multipliers $\left\{j_{\vec{q}}\right\}$ each of which
enforcing the value of $\bar{h}_{\vec{q}}=\langle h_{\vec{q}}\rangle$,
we write the partition function of the surface as
\begin{equation}
Z_G\left[A_p,\left\{j_{\vec{q}}\right\}\right]=
\int D\left[\left\{h_{\vec{q}}\right\}\right]\, 
\exp\left\{-\beta\left[ {\cal H}\left(\left\{j_{\vec{q}}
\right\}\right)
-\sum_{\vec{q}}h_{\vec{q}}j_{\vec{q}}\right]\right\},
\label{partition}
\end{equation}
where $\beta=(k_BT)^{-1}$. The associated Gibbs free energy is
\begin{equation}
G\left[A_p,\left\{j_{\vec{q}}\right\}\right]=-k_BT\ln Z_G.
\label{gibbs}
\end{equation}
From Eqs.(\ref{partition}) and (\ref{gibbs}) it is easy to derive the
following relation
\begin{equation}
\bar{h}_{\vec{q}}=\langle h_{\vec{q}}\rangle=-
\frac{dG}{dj_{\vec{q}}},
\end{equation}
and
\begin{equation}
\langle h_{\vec{q}} h_{-\vec{q}}\rangle - 
\langle h_{\vec{q}}\rangle \langle h_{-\vec{q}}\rangle =
-k_BT \frac{d^2G}{dj_{\vec{q}}dj_{-\vec{q}}}.
\label{hq2}
\end{equation}
The Helmholtz free energy $F$ is related to $G$ via
\begin{equation}
F\left[A_p,\left\{\bar{h}_{\vec{q}}\right\}\right]=
G\left[A_p,\left\{j_{\vec{q}}\right\}\right]+\sum_{\vec{q}}
\bar{h}_{\vec{q}}j_{\vec{q}},
\label{fgrel}
\end{equation}
where
\begin{equation}
\frac{dF}{d\bar{h}_{\vec{q}}}=j_{\vec{q}}.
\label{fgrel2}
\end{equation}
If we use expression (\ref{fourierfenergy}) for the Helmholtz free
energy, we find from Eq.(\ref{fgrel2}) that
\begin{equation}
j_{\vec{q}}=l^2\left[\sigma q^2+\kappa q^4+{\cal O}(q^6)\right] 
\bar{h}_{-\vec{q}}+\cdots
\label{hjrel}
\end{equation}
[note that $\bar{h}_{\vec{q}}\left(j_{\vec{q}}=0\right)=0$]. Combining
Eqs.(\ref{fourierfenergy}), (\ref{fgrel}), and (\ref{hjrel}) we obtain
to the following expression for Gibbs free energy
\begin{equation}
G=F\left(\left\{\bar{h}_{\vec{q}}\right\}=
\left\{0\right\}\right)-\sum_{\vec{q}}
\frac{j_{\vec{q}}j_{-\vec{q}}}{2l^2[\sigma q^2+\kappa q^4+{\cal O}(q^6)]}
+{\cal O} \left(|j_{\vec{q}}|^4\right).
\label{gibbsenergy}
\end{equation}
When this expression for $G$ is substituted in Eq.(\ref{hq2}) and
evaluated for $\{j_{\vec{q}}\}=\{0\}$ (which corresponds to the
reference state $\{\bar{h}_{\vec{q}}\}=\{0\}$), we find that the mean
square amplitude of the fluctuations with a wave-vector $\vec{q}$ (the
``spectral intensity'') is given by
\begin{equation}
\langle h_{\vec{q}}h_{-\vec{q}}\rangle
\Big|_{\{\bar{h}_{\vec{q}}\}=\{0\}}=
\langle |h_{\vec{q}}|^2\rangle\Big|_{\{\bar{h}_{\vec{q}}\}=\{0\}}=
\frac{k_BT}{l^2[\sigma q^2+\kappa q^4+{\cal O}(q^6)]}.
\label{fluctexpress}
\end{equation}
This result, which quantifies the magnitude of the fluctuations around
the flat equilibrium state, provides a second (``fluctuation'') route
for calculating $\sigma$ and $\kappa$ (but not for the saddle-splay
modulus $\bar{\kappa}$). It is frequently quoted in an incorrect form
with $\sigma_0$ and $\kappa_0$, the coefficients in the Helfrich
Hamiltonian (\ref{helfhamiltonian}), instead of $\sigma$ and $\kappa$.
The equivalence of the two routes to membrane elasticity is expressed
by the fact that the elastic coefficients appearing in expression
(\ref{fluctexpress}) are the same as those obtained from
Eqs.(\ref{mechstension})--(\ref{mechgauss}), and which are associated
the ``equilibrium'' route. In the next section we use
Eqs.(\ref{mechstension})--(\ref{mechgauss}) to derive
statistical-mechanical expressions for the elastic coefficients.
Then, in section \ref{numerical}, we demonstrate, using computer
simulations of a bilayer membrane model, the agreement between the two
different methods of calculation.

\section{Thermodynamic expressions for the elastic coefficients}
\label{expressions} 

\subsection{The surface tension}
\label{sectension}

Let us return to the equilibrium route to membrane elasticity and to
expressions (\ref{mechstension})--(\ref{mechgauss}) which describe the
relation between the free energy and the elastic coefficients. The
surface tension can be computed by comparing the free energy of the
membrane at the reference state (which is assumed to be flat) and the
free energy of a flat membrane with a slightly larger area. These two
membranes are shown schematically, without the underlying microscopic
details, in Figs.~\ref{membrane1} (a) and (b). We reemphasize that the
total number of amphiphilic molecules which form the membrane is
fixed, and that the surface tension should be related to the free
energy dependence on the area density of the amphiphiles (rather than
the free energy cost to add molecules to the membrane). The {\em
  characteristic surface}\/ of the membrane to which the free energy
is applied, is chosen as the mid surface between the two layers. The
total volume of the membrane is assumed to be fixed; otherwise, an
additional term involving the volume compression modulus must be
introduced in Eq.(\ref{helffenergy}).

It is important to remember that in Figs.~\ref{membrane1} (a) and (b),
only the mean configurations of the surface (in the reference and
deformed states) are depicted, and that the surface undulates around
these (ensemble) average conformations. In other words, ``the state of
the surface'' refers to its average conformation. As has been
discussed earlier in section \ref{routes}, normal forces must be
applied in order to deform the surface from its reference state
\cite{remarkforces}. If the membrane is embedded in a solution and
placed in a container, than these forces can be generated by deforming
the entire container, as demonstrated in Fig.~\ref{membrane1} (d).
Such a system can be conceptually divided into bulk aqueous phases and
the interface between them which includes the membrane and the
adjacent hydration layers. The volumes of the bulk phases above and
below the membrane are fixed by the presence of solute particles that
cannot permeate the membrane. The deformation of the boundaries of the
container ``percolates'' to the interface and the latter acquires the
shape of the surface of the container.  However, since the bulk
solution is fluid and has a vanishing shear modulus, its deformation
without changing its volume does not add any contribution to the free
energy.

Even thought real bilayer systems are always embedded in a solvent
(which influences their elastic properties), the calculation of the
surface tension can be also performed for model systems that exclude
the latter and leave only the interfacial region. This is possible due
to the fact that the surface tension can be calculated by considering
a deformed {\em flat} membrane. Such a membrane can be uniquely
defined by the perimeter ${\cal P}\left(\bar{h}(\vec{r})\right)$ of
the characteristic surface [represented by open circles in
Fig.~\ref{membrane1} (d)]. The free energy of the membrane can be
derived from the partition function $Z$ via the relation
\begin{equation}
F=-k_BT\ln Z.
\label{fzrel}
\end{equation} 
The expression for the partition function must take into account the
microscopic nature of the membrane, and the potential energy $E$ due to
the interactions between the amphiphilic molecules. In what follows
we assume that $E$ can be written as the sum of pair interactions
between the atoms (``interaction sites'') forming the molecules
\begin{equation}
E=\sum_{\langle \alpha\beta\rangle} \phi\left(r^{\alpha\beta}\right),
\label{potenergy}
\end{equation}
where $r^{\alpha\beta}$ is the distance between atoms $\alpha$ and
$\beta$, and summation over all pairs of atoms $\langle
\alpha\beta\rangle$ is performed. The various interactions are not
identical but rather pair-dependent, as each amphiphilic molecule is
typically composed of many different atoms. They also depend on
whether the atoms belong to different molecules or part of the same
amphiphile. In the latter case some atoms are covalently bonded what
brings in an additional contribution to $E$. For brevity we will omit
the subscripts of the potential and the indices of the argument
$r^{\alpha\beta}$ will serve as an indicator of the specific
potential. With the potential energy described by
Eq.(\ref{potenergy}), the partition function is given by
\begin{equation}
Z=\sum_{{\cal P}{\rm\ -\ Conf.}}
\exp\left(-\sum_{\langle \alpha\beta \rangle}
\phi\left(r^{\alpha\beta}\right)/k_BT\right),
\label{parti1}
\end{equation}
where the sum runs over all the conformations is which the perimeter
of the characteristic surface is depicted by the closed curve ${\cal
  P}$. Our assumption that the membrane has no spontaneous curvature
guarantees that its average conformation is indeed flat.
Alternatively, one may consider the system together with the bulk
phases, and replace the sum in Eq.(\ref{parti1}) with integration of
the coordinates of all atoms $\{\vec{r}^{\, \gamma}\}$ over the entire
volume of container (or the simulation cell) $V_{\rm cell}$
\begin{equation}
Z=\int_{V{\rm cell}}\prod_{\gamma=1}^{N}d\vec{r}^{\, \gamma}
\exp\left(-\sum_{\langle \alpha\beta \rangle}
\phi\left(r^{\alpha\beta}\right)/k_BT\right).
\label{parti2}
\end{equation}
In addition to the above integral, it is necessary to specify the
boundary conditions for the positions of the amphiphiles near the
walls of the container, so that the perimeter of the characteristic
surface would be described by ${\cal P}$.

Let us assume that our cell (container) has a square cross section of
linear size $L_p$ with $-L_p/2\leq x,y< +L_p/2$. The deformation of
the cell depicted in Fig.~\ref{membrane1} (d) can be described by the
following linear transformation
\begin{equation}
\left(\begin{array}{c}
r_x\\ r_y \\ r_z\end{array}\right)=
\left(\begin{array}{ccc}
1 & 0 & 0\\
0 & 1 & 0 \\
\epsilon & 0 & 1  \end{array}\right)
\left(\begin{array}{c}
R_x\\ R_y \\ R_z\end{array}\right),
\label{lintransform}
\end{equation}
which maps every point $\vec{R}$ on the boundaries of the undeformed
cell to its strained spatial position $\vec{r}$. The characteristic
surface has the same shape as the upper and lower faces of the cell,
and its area is given by
\begin{equation}
A=A_p\sqrt{1+\epsilon^2}= 
A_p\left(1+\frac{\epsilon^2}{2}
+{\cal O}\left(\epsilon ^4\right)\right),
\label{stensarea}
\end{equation}
where $A_p=L_p^2$ is the area of the reference surface. Since the
deformed surface which we consider is flat, its free energy is given
by [see Eqs.(\ref{helffenergy}) and (\ref{stensarea})]
\begin{equation}
F=F\left(\epsilon=0\right)+\sigma A_p\frac{\epsilon^2}{2}\cdots\ ,
\end{equation}
from which we conclude that 
\begin{equation}
\sigma=\frac{1}{A_p}\frac{d^2F}{d\epsilon^2}\Biggm|_{\epsilon=0}.
\label{sigmafrel}
\end{equation}
Using the relation between the free energy and the partition function
(\ref{fzrel}), we may also write the above result in the following
form
\begin{equation}
\sigma=-\frac{k_BT}{A_p}
\left[\frac{1}{Z}\frac{d^2Z}{d\epsilon^2}
-\frac{1}{Z^2}\left(\frac{dZ}{d\epsilon}\right)^2\right]\Biggm|_{\epsilon=0}.
\label{surfacetension}
\end{equation}

If we now turn to our expression (\ref{parti2}) for the partition
function, we notice that it depends on $\epsilon$ only through the
integration volume $V_{\rm cell}$. The differentiation of $Z$ with
respect to $\epsilon$, however, could be carried out more easily if
the dependence on $\epsilon$ is removed from $V_{\rm cell}$ and
brought into the integrand. In other words, we wish to change the
integration variables in (\ref{parti2}) from $\vec{r}^{\, \gamma}$ to
$\vec{R}^{\, \gamma}$, where the latter are confined inside the
undeformed cell. This is achieved using transformation
(\ref{lintransform}), which originally described the deformation of
the boundary points, and is now being applied inside the volume of
integration \cite{remarktrans}. With the new set of variables, the distance between two
atoms is given by
\begin{equation}
r^{\alpha\beta}=\left[\left(R^{\alpha\beta}\right)^2+
2\epsilon R_x^{\alpha\beta}R_z^{\alpha\beta}+
\epsilon^2\left(R_x^{\alpha\beta}\right)^2\right]^{1/2}.
\label{rstension}
\end{equation}
In the undeformed reference state
$r^{\alpha\beta}(\epsilon=0)=R^{\alpha\beta}$. The partition function
reads
\begin{equation}
Z=\int_{V_0}\prod_{\gamma=1}^{N}d\vec{R}^{\, \gamma}
\exp\left(-\sum_{\langle \alpha\beta \rangle}
\phi\left[\sqrt{\left(R^{\alpha\beta}\right)^2+
2\epsilon R_x^{\alpha\beta}R_z^{\alpha\beta}+
\epsilon^2\left(R_x^{\alpha\beta}\right)^2}\right]/k_BT\right),
\label{parti3}
\end{equation}
where $V_0\equiv V_{\rm cell}(\epsilon=0)$ is the volume of the
undeformed cell.  The Jacobian of the transformation has been
eliminated from the integrand in the above expression since it is
unity.  The differentiation of $Z$ with respect to $\epsilon$ is now
straightforward but lengthy. We skip the details of the calculation, and
write below the final expressions for the first and second
derivatives, evaluated for $\epsilon=0$ [only the value at
$\epsilon=0$ is required in Eq.(\ref{surfacetension})]
\begin{equation}
\frac{dZ}{d\epsilon}\Biggm|_{\epsilon=0} =
\int_{V_0}\prod_{\gamma=1}^{N}d\vec{R}^{\, \gamma}
\exp\left[-\sum_{\langle \alpha\beta \rangle}
\phi\left(R^{\alpha\beta}\right)\right]
\times \left[-\sum_{\langle\alpha\beta\rangle}
\frac{\phi'\left(R^{\alpha\beta}\right)}{k_BT}
\frac{R_x^{\alpha\beta}R_z^{\alpha\beta}}{R^{\alpha\beta}}\right],
\end{equation}
and
\begin{eqnarray}
&\,&\frac{d^2Z}{d\epsilon^2}\Biggm|_{\epsilon=0}=
\int_{V_0}\prod_{\gamma=1}^{N}d\vec{R}^{\, \gamma}
\exp\left[-\sum_{\langle \alpha\beta \rangle}
\phi\left(R^{\alpha\beta}\right)\right]
\times \left\{\left[\sum_{\langle\alpha\beta\rangle}
\frac{\phi'\left(R^{\alpha\beta}\right)}{k_BT}
\frac{R_x^{\alpha\beta}R_z^{\alpha\beta}}{R^{\alpha\beta}}\right]^2
\right.
\\
&-&\left.\sum_{\langle\alpha\beta\rangle}\left[
\frac{\phi''\left(R^{\alpha\beta}\right)}{k_BT}
\left(\frac{R_x^{\alpha\beta}R_z^{\alpha\beta}}{R^{\alpha\beta}}\right)^2
+ \frac{\phi'\left(R^{\alpha\beta}\right)}{k_BT}
\frac{\left(R_x^{\alpha\beta}\right)^2}{R^{\alpha\beta}}
-\frac{\phi'\left(R^{\alpha\beta}\right)}{k_BT}
\frac{\left(R_x^{\alpha\beta}R_z^{\alpha\beta}\right)^2}
{\left(R^{\alpha\beta}\right)^3\phantom{\biggl)}}\right]\right\},
\nonumber
\end{eqnarray}
where $\phi'\equiv d\phi/dR$ and $\phi''\equiv d^2\phi/dR^2$. When these
expression are substituted into Eq.(\ref{surfacetension}) we readily
find that
\begin{eqnarray}
\sigma&=&\frac{1}{A_pk_BT}\left\{
\left[\left\langle\sum_{\langle\alpha\beta\rangle}
\phi'\left(R^{\alpha\beta}\right)\frac{R_x^{\alpha\beta}R_z^{\alpha\beta}}
{R^{\alpha\beta}}\right\rangle\right]^2-
\left\langle\left[\sum_{\langle\alpha\beta\rangle}
\phi'\left(R^{\alpha\beta}\right)\frac{R_x^{\alpha\beta}R_z^{\alpha\beta}}
{R^{\alpha\beta}}\right]^2\right\rangle\right\}
\nonumber \\ 
\ &+&\frac{1}{A_p}\left\langle\sum_{\langle\alpha\beta\rangle}
\phi''\left(R^{\alpha\beta}\right)
\left(\frac{R_x^{\alpha\beta}R_z^{\alpha\beta}}{R^{\alpha\beta}}\right)^2
\right\rangle+
\frac{1}{A_p}\left\langle\sum_{\langle\alpha\beta\rangle}
\phi'\left(R^{\alpha\beta}\right)
\frac{\left(R_x^{\alpha\beta}\right)^2}{R^{\alpha\beta}}
\right\rangle
\nonumber \\
\ &-&\frac{1}{A_p}
\left\langle\sum_{\langle\alpha\beta\rangle}
\phi'\left(R^{\alpha\beta}\right)
\frac{\left(R_x^{\alpha\beta}R_z^{\alpha\beta}\right)^2}
{\left(R^{\alpha\beta}\right)^3}
\right\rangle,
\label{stension0}
\end{eqnarray}
where the thermal averages are evaluated at the undeformed reference
state of the system ($\epsilon=0$). If the system is macroscopically
invariant with respect to reversal of the sign of the $z$ coordinates
($z\rightarrow -z$), then the first term in the above expression for
$\sigma$ vanishes. If, in addition, the system is invariant with
respect to rotation around the $z$ axis ($x\rightarrow y;\ 
y\rightarrow -x$), then another expression can be derived with
$R_x^{\alpha\beta}$ replaced by $R_y^{\alpha\beta}$. Defining
$R_t^{\alpha\beta}\equiv\sqrt{\left(R_x^{\alpha\beta}\right)^2
  +\left(R_y^{\alpha\beta}\right)^2}$, we finally arrive to the
following expression:
\begin{eqnarray}
\sigma&=&-\frac{1}{2A_pk_BT}
\left\langle\left[\sum_{\langle\alpha\beta\rangle}
\phi'\left(R^{\alpha\beta}\right)\frac{R_x^{\alpha\beta}R_z^{\alpha\beta}}
{R^{\alpha\beta}}\right]^2
+\left[\sum_{\langle\alpha\beta\rangle}
\phi'\left(R^{\alpha\beta}\right)\frac{R_y^{\alpha\beta}R_z^{\alpha\beta}}
{R^{\alpha\beta}}\right]^2
\right\rangle
\nonumber \\ 
\ &+&\frac{1}{2A_p}\left\langle\sum_{\langle\alpha\beta\rangle}
\phi''\left(R^{\alpha\beta}\right)
\left(\frac{R_t^{\alpha\beta}R_z^{\alpha\beta}}{R^{\alpha\beta}}\right)^2
\right\rangle+
\frac{1}{2A_p}\left\langle\sum_{\langle\alpha\beta\rangle}
\phi'\left(R^{\alpha\beta}\right)
\frac{\left(R_t^{\alpha\beta}\right)^2}{R^{\alpha\beta}}
\right\rangle
\nonumber \\
\ &-&\frac{1}{2A_p}
\left\langle\sum_{\langle\alpha\beta\rangle}
\phi'\left(R^{\alpha\beta}\right)
\frac{\left(R_t^{\alpha\beta}R_z^{\alpha\beta}\right)^2}
{\left(R^{\alpha\beta}\right)^3}
\right\rangle.
\label{stension1}
\end{eqnarray}
This expression can be also written in the following compact form
\begin{equation}
\sigma=L_z\left[\frac{C_{xzxz}+C_{yzyz}-P_{xx}-P_{yy}}{2}\right]
\equiv L_z\mu_{zt},
\label{stension1b}
\end{equation}
where $L_z$ is the linear size of the system (the cell) in the $z$
direction (normal to the membrane), while $P$ and $C$ denote the
pressure tensor and the tensor of elastic constants of the system. The
quantity $\mu_{zt}$ is the shear modulus associated with the
deformation depicted at Fig.~\ref{membrane1} (d) \cite{farago,wallace}.
 
In is interesting to compare the above results
(\ref{stension1})--(\ref{stension1b}) with the much better known (and
more frequently used) expression for the surface tension
\cite{rowlinson_widom,rowlinson}
\begin{equation}
\tilde{\sigma}=
\frac{1}{2A_p}\left\langle\sum_{\langle\alpha\beta\rangle}
\phi'\left(R^{\alpha\beta}\right)
\frac{\left(R_t^{\alpha\beta}\right)^2-2\left(R_z^{\alpha\beta}\right)^2}
{R^{\alpha\beta}}
\right\rangle=L_z\left[\frac{2P_{zz}-P_{xx}-P_{yy}}{2}\right].
\label{stension2}
\end{equation} 
The latter expression is obtained when one considers the variation of
the free energy resulting from the (volume-preserving) variation of
the {\em projected area}\/ $A_p$
\begin{equation}
\tilde{\sigma}=\frac{\partial F}{\partial A_p}\Biggm|_{V}.
\end{equation}
The deformed state associated with the surface tension
$\tilde{\sigma}$ is shown in Fig.~\ref{membrane1} (c). For fluid
membranes we anticipate that $\sigma=\tilde{\sigma}$ since the
difference between them
\begin{equation}
\sigma-\tilde{\sigma}= L_z\left[\frac{C_{xzxz}+C_{yzyz}}{2}-P_{zz}\right]
\equiv L_z\mu_{tz},
\end{equation}
is proportional to the   shear modulus $\mu_{tz}$ associated  with the
deformation  shown  in   Fig.~\ref{membrane1} (e).  The  shear modulus
$\mu_{tz}$  is  expected   to  vanish  because   the   areas  of   the
characteristic surfaces of the  membranes in Figs.~\ref{membrane1} (a)
and (e) are identical; and the Helfrich free energy of a flat membrane
depends only on the area of the characteristic surface, but not on the
orientation of the plane of the membrane with  respect to the walls of
the  container. This argument   for the  coincidence  of  $\sigma$ and
$\tilde{\sigma}$   could be  applied directly   to   the membranes  in
Figs.~\ref{membrane1} (b) and (c), whose characteristic areas (as well
as their volumes)  are also identical. The   tilt of the cell's  wall,
however, can be safely ignored  only in the thermodynamic limit,  when
the   width  of the  membrane becomes  much  smaller than  its lateral
dimensions. If the system is not sufficiently  large than the Helfrich
form for the free energy in which the membrane is associated with a 2D
characteristic surface is not entirely applicable. The finite width of
the membrane must show up  in the expression  for the free energy, and
the surface tensions   $\sigma$ and $\tilde{\sigma}$  do not perfectly
agree.

\subsection{The bending modulus}
\label{secbending}

The bending modulus can be calculated by considering a deformation of
the characteristic surface from a flat to cylindrical geometry. The
deformation, which is depicted in Fig.~\ref{membrane2}, can be
described by the following nonlinear transformation of the boundaries
of the cell [compare with Eq.(\ref{lintransform})]
\begin{eqnarray}
r_x &=& R_x
\nonumber \\
r_y &=& R_y
\nonumber \\
r_z &=& R_z+\sqrt{R_0^2-x^2}-\sqrt{R_0^2-L_p^2/4},
\label{nonlintransform}
\end{eqnarray}
where $-L_p/2\leq x< +L_p/2$, and $R_0\gg L_p$ is the radius of
curvature of the cylinder. The integrated total curvature
[Eq.(\ref{intcurv})] of the characteristic surface is
\begin{equation}
\bar{J}=\frac{L_p}{R_0},
\end{equation}
and its area is
\begin{equation}
A=A_p+2\arcsin\left(\frac{L_p}{2R_0}\right)\simeq 
A_p\left[1+\frac{1}{24}{\bar{J}}^{\,2}+{\cal O}({\bar{J}}^{\,4})\right].
\end{equation}
The free energy is, hence, given by
\begin{equation}
F=\left[\frac{\sigma L^2}{24}+\frac{1}{2}\kappa\right]{\bar{J}}^{\,2}+\cdots ,
\end{equation}
from which we deduce the following relation
\begin{equation}
\frac{\sigma L^2}{12}+\kappa =\frac{1}{A_p}\frac{d^2F}{dJ^2}\Biggm|_{J=0},
\label{kappafrel}
\end{equation}
where
\begin{equation}
J\equiv\frac{1}{R_0}.
\end{equation}

The calculation of the r.h.s of the above equation is very similar to
the one presented in section \ref{sectension} which was based on
expression (\ref{parti3}) for the partition function. The deformed
pair distance, which in that case was given by Eq.(\ref{rstension}),
is now depicted by the following relation
\begin{equation}
r^{\alpha\beta}=\left[\left(R^{\alpha\beta}\right)^2-
2\bar{X}^{\alpha\beta}R_x^{\alpha\beta}R_z^{\alpha\beta}J+
\left(\bar{X}^{\alpha\beta}\right)^2\left(R_x^{\alpha\beta}\right)^2J^2
\right]^{1/2},
\label{rbending}
\end{equation}
where 
\begin{equation}
\bar{X}^{\alpha\beta}\equiv \frac{X^{\alpha}+X^{\beta}}{2}
\end{equation}
is the average of the $x$-coordinates of atoms $\alpha$ and $\beta$
(in the undeformed state). Comparing Eqs.(\ref{rstension}) and
(\ref{rbending}), and respectively, Eqs.(\ref{sigmafrel}) and
(\ref{kappafrel}), it is easy to realize that the result of the
calculation is the following expression
\begin{eqnarray}
\frac{\sigma L^2}{12}+\kappa&=&\frac{1}{A_pk_BT}\left\{
\left[\left\langle\sum_{\langle\alpha\beta\rangle}
\phi'\left(R^{\alpha\beta}\right)\frac{\bar{X}^{\alpha\beta}
R_x^{\alpha\beta}R_z^{\alpha\beta}}
{R^{\alpha\beta}}\right\rangle\right]^2-
\left\langle\left[\sum_{\langle\alpha\beta\rangle}
\phi'\left(R^{\alpha\beta}\right)\frac{\bar{X}^{\alpha\beta}
R_x^{\alpha\beta}R_z^{\alpha\beta}}
{R^{\alpha\beta}}\right]^2\right\rangle\right\}
\nonumber \\ 
\ &+&\frac{1}{A_p}\left\langle\sum_{\langle\alpha\beta\rangle}
\phi''\left(R^{\alpha\beta}\right)
\left(\frac{\bar{X}^{\alpha\beta}
R_x^{\alpha\beta}R_z^{\alpha\beta}}{R^{\alpha\beta}}\right)^2
\right\rangle+
\frac{1}{A_p}\left\langle\sum_{\langle\alpha\beta\rangle}
\phi'\left(R^{\alpha\beta}\right)
\frac{\left(\bar{X}^{\alpha\beta}R_x^{\alpha\beta}\right)^2}{R^{\alpha\beta}}
\right\rangle
\nonumber \\
\ &-&\frac{1}{A_p}
\left\langle\sum_{\langle\alpha\beta\rangle}
\phi'\left(R^{\alpha\beta}\right)
\frac{\left(\bar{X}^{\alpha\beta}R_x^{\alpha\beta}R_z^{\alpha\beta}\right)^2}
{\left(R^{\alpha\beta}\right)^3}
\right\rangle,
\label{bending0}
\end{eqnarray}
which is similar to Eq.(\ref{stension0}), except for the fact that
$R^{\alpha\beta}_x$ is everywhere replaced by
$-\bar{X}^{\alpha\beta}R^{\alpha\beta}_x$. 

Among the five terms on the r.h.s of Eq.(\ref{bending0}), only the
second involves averages of quantities including the product
$\bar{X}^{\alpha\beta}\bar{X}^{\gamma\delta}$ with
$\langle\alpha\beta\rangle\neq\langle\gamma\delta\rangle$. In the
other four terms, the quantities $\bar{X}^{\alpha\beta}$ and
$\left(\bar{X}^{\alpha\beta}\right)^2$ can be replaced by their
averages
\begin{equation}
\left\langle\bar{X}^{\alpha\beta}\right\rangle=\frac{1}{L_p}
\int_{-L_p/2}^{L_p/2}x\,dx=0,
\end{equation}
and
\begin{equation}
\left\langle\left(\bar{X}^{\alpha\beta}\right)^2\right\rangle=
\frac{1}{L_p}\int_{-L_p/2}^{L_p/2}
x^2\,dx=\frac{L_p^2}{12},
\label{l212}
\end{equation}
since they multiply quantities which depend only on the separation
between atoms $\alpha$ and $\beta$ and whose averages, therefore, are
independent of the location of the pair (provided the system is
invariant to translations in the $x$ and $y$ directions). This, in
combination with Eq.(\ref{stension0}), yield the following expression
for $\kappa$
\begin{equation}
\kappa=\frac{1}{A_pk_BT}\left\{
\left\langle\frac{L_p^2}{12}\left[\sum_{\langle\alpha\beta\rangle}
\phi'\left(R^{\alpha\beta}\right)\frac{R_x^{\alpha\beta}R_z^{\alpha\beta}}
{R^{\alpha\beta}}\right]^2\right\rangle
-\left\langle\left[\sum_{\langle\alpha\beta\rangle}
\phi'\left(R^{\alpha\beta}\right)\frac{\bar{X}^{\alpha\beta}
R_x^{\alpha\beta}R_z^{\alpha\beta}}
{R^{\alpha\beta}}\right]^2\right\rangle\right\}.
\end{equation}
Replacing $R_x^{\alpha\beta}$ with $R_y^{\alpha\beta}$, and
$\bar{X}^{\alpha\beta}$ with
$\bar{Y}^{\alpha\beta}\equiv(Y^{\alpha}+Y^{\beta})/2$, we obtain the
``symmetric'' formula
\begin{eqnarray}
\kappa&=&\frac{1}{2A_pk_BT}\left\{
\left\langle\frac{L_p^2}{12}\left[\sum_{\langle\alpha\beta\rangle}
\phi'\left(R^{\alpha\beta}\right)\frac{R_x^{\alpha\beta}R_z^{\alpha\beta}}
{R^{\alpha\beta}}\right]^2
+\frac{L_p^2}{12}\left[\sum_{\langle\alpha\beta\rangle}
\phi'\left(R^{\alpha\beta}\right)\frac{R_y^{\alpha\beta}R_z^{\alpha\beta}}
{R^{\alpha\beta}}\right]^2\right\rangle\right.
\nonumber \\
&-&\left.\left\langle\left[\sum_{\langle\alpha\beta\rangle}
\phi'\left(R^{\alpha\beta}\right)\frac{\bar{X}^{\alpha\beta}
R_x^{\alpha\beta}R_z^{\alpha\beta}}
{R^{\alpha\beta}}\right]^2\right\rangle
-\left\langle\left[\sum_{\langle\alpha\beta\rangle}
\phi'\left(R^{\alpha\beta}\right)\frac{\bar{Y}^{\alpha\beta}
R_y^{\alpha\beta}R_z^{\alpha\beta}}
{R^{\alpha\beta}}\right]^2\right\rangle
\right\}.
\label{bending1}
\end{eqnarray}

It is important to remember here that the above expression for
$\kappa$ (\ref{bending1}) applies to square membranes only with the
origin of axes located at the center of the membrane so that
$-L_p/2\leq x,y< +L_p/2$. A formula which does not depend neither
on the shape of the membrane nor on the location of the origin is
obtained as follows: The first and third terms in Eq.(\ref{bending1})
can be written jointly in the following form
\begin{eqnarray}
\sum_{\langle\alpha\beta\rangle}\sum_{\langle\gamma\delta\rangle}
\left\langle
\phi'\left(R^{\alpha\beta}\right)\phi'\left(R^{\gamma\delta}\right)
\frac{R_x^{\alpha\beta}R_z^{\alpha\beta}R_x^{\gamma\delta}R_z^{\gamma\delta}}
{R^{\alpha\beta}R^{\gamma\delta}}
\left(\frac{L^2}{12}-\bar{X}^{\alpha\beta}\bar{X}^{\gamma\delta}\right)
\right\rangle=
\\
\sum_{\langle\alpha\beta\rangle}\sum_{\langle\gamma\delta\rangle}
\left\langle
\phi'\left(R^{\alpha\beta}\right)\phi'\left(R^{\gamma\delta}\right)
\frac{R_x^{\alpha\beta}R_z^{\alpha\beta}R_x^{\gamma\delta}R_z^{\gamma\delta}}
{R^{\alpha\beta}R^{\gamma\delta}}
\left[\frac{L^2}{12}-\left(\bar{X}^{\alpha\beta,\gamma\delta}\right)^2+
\left(\Delta_X^{\alpha\beta,\gamma\delta}\right)^2\right]
\right\rangle,
\label{bending2}
\end{eqnarray}
where 
\begin{equation}
\bar{X}^{\alpha\beta,\gamma\delta}\equiv\frac{\bar{X}^{\alpha\beta}
+\bar{X}^{\gamma\delta}}{2},
\end{equation}
and
\begin{equation}
\Delta_X^{\alpha\beta,\gamma\delta}\equiv\frac{\bar{X}^{\alpha\beta}
-\bar{X}^{\gamma\delta}}{2}.
\end{equation}
The terms appearing before the square brackets in Eq.(\ref{bending2})
depend only on the relative coordinates of atoms with respect to each
other. Therefore, the average of
$\left(\bar{X}^{\alpha\beta,\gamma\delta}\right)^2$ (the second term
in square brackets, which depends only the location of the center of
the pair/triplet/quartet in question) can be performed separately. As
in Eq.(\ref{l212}) we have $\left\langle
  \left(\bar{X}^{\alpha\beta,\gamma\delta}\right)^2\right\rangle
=L_p^2/12$, what leads to the cancellation of the first two terms in
square brackets in Eq.(\ref{bending2}). Applying the same argument for
the second and fourth terms in Eq.(\ref{bending1}), and defining
$\bar{Y}^{\alpha\beta,\gamma\delta}\equiv\left(\bar{Y}^{\alpha\beta}
  +\bar{Y}^{\gamma\delta}\right)/2$, and
$\Delta_Y^{\alpha\beta,\gamma\delta}\equiv\left(\bar{Y}^{\alpha\beta}
  -\bar{Y}^{\gamma\delta}\right)/2$, we arrive to the following
expression
\begin{equation}
\kappa=\frac{1}{2A_pk_BT}
\sum_{\langle\alpha\beta\rangle}\sum_{\langle\gamma\delta\rangle}
\left\langle
\phi'\left(R^{\alpha\beta}\right)\phi'\left(R^{\gamma\delta}\right)
\frac{R_z^{\alpha\beta}R_z^{\gamma\delta}}
{R^{\alpha\beta}R^{\gamma\delta}}
\left[
R_x^{\alpha\beta}R_x^{\gamma\delta}
\left(\Delta_X^{\alpha\beta,\gamma\delta}\right)^2
+R_y^{\alpha\beta}R_y^{\gamma\delta}
\left(\Delta_Y^{\alpha\beta,\gamma\delta}\right)^2
\right]
\right\rangle,
\label{bending3}
\end{equation}
which is the more general form for expression (\ref{bending1}) since
it is independent of the shape of the membrane and of the location of
the origin of axes.
 
The deformed membrane, shown schematically in gray shade in
Fig.~\ref{membrane2}, may be considered as part of a closed
cylindrical vesicle (depicted by the dashed line
Fig.~\ref{membrane2}). Accordingly, one may argue that its free energy
is given by
\begin{equation}
F=\frac{\theta}{2\pi}F_{\rm vesicle}
\end{equation}
where $F_{\rm vesicle}$ is the free energy of the vesicle and $\theta$
is the apex angle of the deformed membrane. This relation, however, is
incorrect since $F_{\rm vesicle}$ includes a term which is unique to
closed vesicles and should be omitted in the case of open membranes.
The additional contribution to $F_{\rm vesicle}$ which has been termed
``the {\em area-difference}\/ elastic energy'', should not be confused
with the bending energy. The latter is the free energy required to
bend the membrane while keeping its area density fixed.  The former,
on the other hand, originates from the simple fact that upon closure
of the vesicle, it becomes impossible to preserve the area densities
of the amphiphiles in both the outer and the inner monolayers. The
outer monolayer is stretched and the inner monolayer is compressed
relative to the mid characteristic surface. The elastic energy
resulting from such curvature-induced changes in the monolayer areas
is a {\em non-local}\/ effect because the monolayers are capable of
independent lateral redistribution to equalize the area per molecule
of each leaflet. The distinction between (local) bending elasticity
and (non-local) area-difference elasticity has been discussed by
Helfrich, not long after introducing his famous Hamiltonian
\cite{helfrich2}. The idea, however, did not gain much popularity
until the issue has been analyzed systematically by Svetina et
al.~some years later \cite{svetina}.  Early theoretical works and
experimental measurements of the bending modulus failed to separate
the local and non-local contributions \cite{remarkade}. This is not
the case with our expression (\ref{bending3}) for $\kappa$ which has
been derived by considering an open membrane. For an open membrane,
the two leaflets have the same area as the top (button) surface of the
containers and, consequently, area-difference elasticity do not show
up.

\subsection{The saddle-splay modulus}

Finally, we derive our expression for the saddle-splay modulus
$\bar{\kappa}$. The following transformation, applied to the boundaries
of the container
\begin{eqnarray}
r_x &=& R_x
\nonumber \\
r_y &=& R_y
\nonumber \\
r_z &=& R_z+\sqrt{R_0^2-x^2-y^2}-\sqrt{R_0^2-L_p^2/2}
\label{nonlintransform2}
\end{eqnarray}
(with $-L_p/2\leq x< +L_p/2$), describes a deformation of the surface
to spherical geometry where the sphere's radius $R_0\gg L_p$. It is
not difficult to show that the free energy of the spherical surface is
given by
\begin{equation}
F=A_p\left[\frac{\sigma L^2}{12}+2\kappa+\bar{\kappa}\right]H^2+\cdots\ ,
\end{equation}
where $H=1/R_0$. From the above expression for $F$, the following
relation
\begin{equation}
\frac{\sigma L^2}{6}+4\kappa+2\bar{\kappa}=
\frac{1}{A_p}\frac{d^2F}{dH^2}\Biggm|_{H=0},
\label{gaussfrel}
\end{equation}
is easily derived. The deformed pair distance is
\begin{equation}
r^{\alpha\beta}=\left[\left(R^{\alpha\beta}\right)^2-
2\left(\bar{X}^{\alpha\beta}R_x^{\alpha\beta}+
\bar{Y}^{\alpha\beta}R_y^{\alpha\beta}\right)R_z^{\alpha\beta}H+
\left(\bar{X}^{\alpha\beta}R_x^{\alpha\beta}+
\bar{Y}^{\alpha\beta}R_y^{\alpha\beta}\right)^2H^2
\right]^{1/2},
\label{rgauss}
\end{equation}
where $\bar{X}^{\alpha\beta}$ and $\bar{Y}^{\alpha\beta}$ have been
defined in section \ref{secbending}. Since Eqs.(\ref{gaussfrel}) and
(\ref{rgauss}) have, respectively, the same form as
Eqs.(\ref{kappafrel}) and (\ref{rbending}), we immediately conclude
that the r.h.s of Eq.(\ref{gaussfrel}) is given by expression similar
to (\ref{bending0}) in which $\bar{X}^{\alpha\beta}R^{\alpha\beta}_x$
is everywhere exchanged with $\bar{X}^{\alpha\beta}R_x^{\alpha\beta}+
\bar{Y}^{\alpha\beta}R_y^{\alpha\beta}$. Following the same steps
described in the derivation of Eq.(\ref{bending1}) from
(\ref{bending0}), and using the additional relation
\begin{equation}
\left\langle\bar{X}^{\alpha\beta}\bar{Y}^{\alpha\beta}\right\rangle=
\frac{1}{L_p^2}\int_{-L_p/2}^{L_p/2}\int_{-L_p/2}^{L_p/2}
xy\,dxdy=0,
\end{equation}
we finally arrive to the following result
\begin{equation}
\bar{\kappa}=-\kappa-\frac{1}{A_pk_BT}\left\langle
\left[\sum_{\langle\alpha\beta\rangle}
\phi'\left(R^{\alpha\beta}\right)\frac{\bar{X}^{\alpha\beta}
R_x^{\alpha\beta}R_z^{\alpha\beta}}
{R^{\alpha\beta}}\right]
\left[\sum_{\langle\alpha\beta\rangle}
\phi'\left(R^{\alpha\beta}\right)\frac{\bar{Y}^{\alpha\beta}
R_y^{\alpha\beta}R_z^{\alpha\beta}}
{R^{\alpha\beta}}\right]\right\rangle.
\label{gauss1}
\end{equation}
This expression applies to square membranes only, with the origin
located at the center of the membrane. The more general expression is
\begin{equation}
\bar{\kappa}=-\kappa-\frac{1}{A_pk_BT}
\sum_{\langle\alpha\beta\rangle}\sum_{\langle\gamma\delta\rangle}
\left\langle
\phi'\left(R^{\alpha\beta}\right)\phi'\left(R^{\gamma\delta}\right)
\frac{R_z^{\alpha\beta}R_z^{\gamma\delta}}
{R^{\alpha\beta}R^{\gamma\delta}}
R_x^{\alpha\beta}\Delta_X^{\alpha\beta,\gamma\delta}
R_y^{\gamma\delta}\Delta_Y^{\alpha\beta,\gamma\delta}
\right\rangle.
\label{gauss2}
\end{equation}

\section{Numerical results}
\label{numerical}

The purpose of the MC simulations which we conducted and present in
this section is twofold: The first is to test the validity and
accuracy of our expressions for the elastic coefficients. The second
is to examine the agreement between the mechanical and the fluctuation
routes to membrane elasticity, as discussed in section \ref{routes}.
The model system whose elastic properties were studied by the
simulations has been described in great details in
Ref.~\cite{farago2}. Briefly, the ``lipids'' that serve as the
building blocks of the membrane consist of three spherical atoms of
diameter $a$ (see Fig.~\ref{lipid}) interacting with each other via
pair-wise LJ potentials (whose details can be found in
Ref.~\cite{farago2}). To avoid the complications involved with
long-range interactions, the LJ potentials have been truncated at some
cut-off separation $R^{\alpha\beta}=r_c=2.5a$ and, in addition,
modified to ensure the vanishing of $\phi$ and its first two
derivatives, $\phi'$ and $\phi''$, at $r_c$. The continuity of the
second derivative of the pair potentials is an important feature since
$\phi''$ appears in our expressions (\ref{stension1}) for $\sigma$.
Two changes have been made in comparison to the original model
presented in Ref.~\cite{farago2}. The first is a small reduction of
the temperature which, in this paper, has been set to $0.9T_0$ where
$T_0$ is the original temperature (in Ref.~\cite{farago2}). The second
is the addition of new interactions between atoms which are part of
the {\em same}\/ molecule. In Ref.~\cite{farago2} the molecule were
linear rigid trimers with a fixed distance $a$ between the centers of
the constituent atoms. Here, we allow some little variations of the
separation between the atoms. The mid atom (labeled 2) has been linked
to the two end atoms (labeled 1 and 3) via harmonic springs with
spring constant $K$ and equilibrium length $a$:
\begin{equation}
\phi(R)=\frac{1}{2}K(R-a)^2,
\label{spring1}
\end{equation}
while the pair potential between the end atoms has been set to
\begin{equation}
\phi(R)=\frac{1}{2}K(R-2a)^2.
\label{spring2}
\end{equation}
We use a large value for the spring constant $K=8000\, k_BT/a^2$, for
which the separations between the atoms do not exceed the order of
$1\%$ of their equlibrium values. While this means that the molecules
in our model are ``almost'' linear and rigid, the use of the above
potentials (\ref{spring1}) and (\ref{spring2}) creates a situation in
which all inter-atomic interactions (whether between atoms belonging
to the same or different molecules) are depicted by smooth potentials;
and so, our expressions for the elastic constants can be used without
any further complications. The total number of lipids in our
simulations was $N=1000$ (500 lipids in each monolayer), and no
additional solvent molecules were included inside the simulation cell
(as if the membrane is vacuum). Periodic boundary conditions were
applied in the plane of the membrane, and no boundaries for the
simulation cell were defined in the normal direction. The linear size
of the (square) membrane was set to $L_p=29.375a$. Subsequent MC
configurations were generated by two types of move attempts:
translations of lipids (which also included some minute changes in the
relative locations of the three atoms with respect to each other) and
rotations around the mid atom. A set of $2N=2000$ move attempts of
randomly chosen molecules is defined as the MC time unit. Both types
of moves (translations and rotations) were attempted with equal
probability, and the acceptance probabilities of both of them was
approximately half. The MC relaxation time has been evaluated in
Ref.~\cite{farago2}. It is of the order of $10^4$ MC time units and
has been very little affected by the changes introduced in the model.
A typical equilibrium configuration of the membrane is shown in
Fig.~\ref{membranefig}.

\subsection{The fluctuation route}

The fluctuation approach for determining the surface tension $\sigma$
and the bending modulus $\kappa$ is straightforward to implement: The profile
of the membrane in our simulations was defined by mapping the system
onto an $8\times8$ grid, and defining the height
$h(\left\{\vec{r}_g\right\})$ of the membrane in each grid cell as the
average of the local heights of the two monolayers. The latter were
evaluated by the mean height of the lipids (whose positions were
identified with the coordinates of their mid atoms) belonging to each
layer, which were instantaneously located inside the local grid cell.
Note that the mesh size $l=L_p/8\simeq 3.67a$ is somewhat larger than
the size of the lipids, as required in our discussion in section
\ref{routes}. The Fourier transform of $h(\left\{\vec{r}_g\right\})$
was obtained using Eq.(\ref{invtransform2}), and the mean squared
amplitudes of the different modes were, eventually, fitted to the
inverse form of Eq.(\ref{fluctexpress})
\begin{equation}
\frac{1}{\langle |h_{\vec{q}}|^2\rangle}=
\frac{l^2[\sigma q^2+\kappa q^4+{\cal O}(q^6)]}{k_BT}.
\label{fluctexpress2}
\end{equation}
The results of this spectral analysis are summarized in
Fig.~\ref{spectrum}, where we plot the value of $1/l^2\langle
|h_{\vec{q}}|^2\rangle$ as a function of $q^2$. The error bars
represent one standard deviation in the estimates of the averages,
which were obtained from simulations of 16 different membranes and a
total number of $1.25\times10^4$ measurements of the spectrum per
membrane. The measurements were done at time intervals of 100 MC time
units. The curve depicts the best fit to Eq.(\ref{fluctexpress2}),
which is obtained when $\sigma$ and $\kappa$ take the following
values:
\begin{eqnarray}
\sigma&=&\left(-0.6\pm 0.2\right)\ \frac{k_BT}{a^2}
\nonumber\\
\kappa&=&\left(46\pm 2\right)\ k_BT.
\label{fluctres}
\end{eqnarray}
The contribution of the $q^6$ term to the fit was, indeed,
significantly smaller than that of the other two terms on the r.h.s.
of Eq.(\ref{fluctexpress2}).

\subsection{The equilibrium route}

While the measurement of $\sigma$ and $\kappa$ using the fluctuation
approach was a relatively straightforward matter, the application of
the equilibrium approach emerged as somewhat more challenging task.
The most significant differences between the two approaches was the
amount of computer resources required for an accurate determination of
the elastic coefficients. The results which we present in this section
have been obtained using 64 nodes on a Beowolf cluster consisting of
Intel architecture PCs, where the CPU time per node was of the order
of three months. The need of such a large computer time should be
compared to the relative ease with which the results in
Eq.(\ref{fluctres}) have been obtained - using a total number of only
16 nodes over a period of about 10 days. The reason that the
equilibrium approach is so much computer-time-consuming is the
``noisy'' nature of the statistics of the terms whose averages are
evaluated in expressions (\ref{stension1}) and (\ref{bending1}). From
the conceptual point of view, the determination of the surface tension
$\sigma$ using expression (\ref{stension1}) is pretty simple. The
determination of the surface tension $\tilde{\sigma}$ from expression
(\ref{stension2}) is even easier since it is a much less noisy
quantity. In fact, the computational effort required for an accurate
determination of the value of $\tilde{\sigma}$ is even smaller than
the one required for the calculation of $\sigma$ by the fluctuation
method. The surface tension $\tilde{\sigma}$ does not apply directly
to membranes with a fixed projected area. Yet, it is expected to
coincide with $\sigma$ in the thermodynamic limit

The determination of $\kappa$ is more complicated. Here we can, in
principle, choose between expressions (\ref{bending1}) and
(\ref{bending3}). The latter is more general (since it is not
restricted to square membranes), but prohibitively time consuming.
This can be understood by considering the number of operations
required for a single measurement of the quantities of interest.
Assuming each atom in our simulations interact with a finite number of
other atoms, the total number of operations required by expression
(\ref{bending3}) is ${\cal O}(N^2)$, while the number required by
expression (\ref{bending1}) is only ${\cal O}(N)$. In our simulations
the total number of atoms is $3000$, which means a difference of about
4 orders of magnitude in efficiency. Using expression (\ref{bending1})
to measure $\kappa$ is, however, tricky because this expression
involves not only the relative locations of the particles with respect
to each other (as in the case of the expressions for the surface
tension), but also the absolute coordinates of atoms. This would {\em
  not}\/ create a problem if only the central coordinates
($\bar{X}^{\alpha\beta}$ and $\bar{Y}^{\alpha\beta}$) of the pairs had
to be found [as one may, naively, conclude from Eq.(\ref{bending1})],
since that among the set including the pair $(\alpha,\beta)$ and all
its periodic images, only one satisfies the requirement $-L_p/2\leq
\bar{X}^{\alpha\beta},\ \bar{Y}^{\alpha\beta}<+L_p/2$. However [and
this becomes clear from the derivation of expression (\ref{bending3})
from expression (\ref{bending1})], what we actually have here is a
periodic boundary conditions problem where the {\em pairs}\/ play the
role of the particles, and $\bar{X}^{\alpha\beta}$ and
$\bar{Y}^{\alpha\beta}$ serve as the coordinates of these
``particles''. This means that each {\em quartet}\/
$\left((\alpha,\beta),(\gamma,\delta)\right)$ is identified as the
pair $(\alpha,\beta)$ and the pair $(\gamma,\delta)$ or its image
nearest to $(\alpha,\beta)$ and, in addition, that the center of the
quartet must satisfy $-L_p/2\leq
\bar{X}^{\alpha\beta,\gamma\delta},\bar{Y}^{\alpha\beta,\gamma\delta}<L_p/2$.
The fact that sometimes a pair must be replaced by one of its images
(which are located outside the boundaries of the simulation cell) is
problematic since this means that the location of the pair, which is
needed in expression (\ref{bending1}), cannot be specified by a single
value. A solution to this problem is obtained by dividing the
simulation cell into stripes parallel to either the $x$ or the $y$
axes [depending on whether we calculate the third or fourth term in
Eq.(\ref{bending1})], and to split the summation over all the pairs to
several partial sums over the pairs included in the different stripes.
The partial sums corresponding to the images of each stripe (which
consist of all the images of the pairs included in the stripe) can be
found with almost no additional effort. The product of two partial
sums gives the contribution of all the quartets consisting of pairs
located inside the two relevant stripes. Depending on the distance
between the stripes (along the relevant axis) and their locations with
respect to the center of the cell, it is {\em usually}\/ easy to
decide in which case a stripe should be replaced by one of its images.
Ambiguities about the correct decision occur in a finite number of
cases (i.e., for a finite number of pairs of stripes). In these cases,
individual decisions must be made for each quartet. The number of such
quartets can be reduced significantly if the system is divided into a
large number of stripes $N_s$, since the narrower the stripes the
smaller the number of pairs included in each one of them. A more
elegant solution is to choose a certain convention about the ways the
contribution from the ambiguous quartets is added to
Eq.(\ref{bending1}). This will inevitably introduce a systematic error
to our estimates of $\kappa$. However, if we make a set of estimates
based on increasingly larger values of $N_s$, we can obtain the
correct averages by extrapolating our results to the limit
$1/N_s\rightarrow 0$. The method, which is described in more details
in the Appendix, can be generalized to handle correctly the
calculation of $\bar{\kappa}$. However, because of the mixing of the
$x$ and $y$ coordinates in Eq.(\ref{gauss1}), the implementation of
the method becomes more complicated.  For this reason, and due to the
fact that the fluctuation approach does not provide a value of
saddle-splay modulus to compare with, we did not use our simulations
to calculate $\bar{\kappa}$.

In section \ref{routes} we have explained in great details why the
elastic coefficients obtained from the fluctuation approach are the
free energy coefficients $\sigma$ and $\kappa$ rather than the
Hamiltonian coefficients $\sigma_0$ and $\kappa_0$. This means that
the quantities in expressions (\ref{stension1}) and (\ref{bending1})
should be averaged over the ensemble of all possible microscopic
configurations. However, it is also easy to understand that the {\em
  same}\/ expressions can be used to calculate the Hamiltonian
coefficients. The latter, which characterize the energy changes caused
by deformations of the flat membrane, can be obtained by restricting
the averages to conformations where $h(\vec{r}_g)=0$ for every grid
cell, thus avoiding the entropic contribution of the thermal
fluctuation to the free energy.  To sample this configuration phase
space one need to accompany every MC move attempt with one or two
(depending on whether the molecule leaves the grid cell or not)
additional moves of adjacent molecules.  Moreover, one can also sample
the phase-space consisting of only those conformations of the membrane
with wave vectors in the range $2\pi/L_p\leq\Lambda<q$. The results of
such a calculation are the wave-dependent coefficients
$\sigma(\Lambda)$ and $\kappa(\Lambda)$. One of the problems which can
be studied by such investigation is the value of the numerical
coefficient $c$ in the formula for the renormalized bending modulus
\cite{peliti_leibler,helfrich3,forster,kleinert}:
\begin{equation}
\kappa(\Lambda)=\kappa_0+c\frac{k_BT}{4\pi}\ln(\Lambda l).
\label{renormalized}
\end{equation}
This problem aroused a renewed interest recently since its has been
suggested that the value of $c$ may be positive, which means (quite
remarkably) that the fluctuations stiffen rather then soften the
membrane \cite{helfrich4,pinnow,nishiyama}. 

While determining the value of $c$ was not possible with our
computer resources, we did use Eq.(\ref{renormalized}) in our analysis
of the results. Our need of Eq.(\ref{renormalized}) and the link that
it provides between $\kappa$ and $\kappa_0$ is related to the peculiar
nature of our simulations which are made in a ``solvent-free''
environment. As has been discussed in section \ref{expressions}, our
expressions for the elastic coefficients have been derived based on
the assumption that the membrane is embedded in solvent and that the
entire container is deformed. In our simulations, however, we have no
container (there are no boundaries for the simulation cell in the $z$
direction) and, so, the applicability of our approach should be
examined carefully. The arguments which we presented in section
\ref{sectension} [see, in particular, the discussion around
Eq.(\ref{parti1})] demonstrate that the presence of solvent is
essential only for the calculation of $\kappa$ and $\bar{\kappa}$, but
not for the calculation of the surface tensions $\sigma$ and
$\tilde{\sigma}$. By contrast, the Hamiltonian coefficients can be
{\em all}\/ measured in a ``solvent-free'' model since they are
extracted from simulations of flat, non-fluctuating, membranes. The
value of $\kappa_0$ and the relation given by Eq.(\ref{renormalized})
provide then an estimate for the value of $\kappa$. Since the
finite-size correction to the value of $\kappa$ grows only
logarithmically with the size of the system, and since $\kappa_0\gg
k_BT$, the difference between $\kappa_0$ and $\kappa$ is not
significantly large. In our simulations it actually falls within the
uncertainty in our estimates of the bending modulus, which means that
$\kappa$ and $\kappa_0$ are practically indistinguishable. In addition
to our measurement of $\kappa_0$, we also measured $\kappa$ directly
from the simulations. As we have just explained above, such a
measurement is expected to fail and to lead to the incorrect
conclusion that $\kappa=0$. We used this incorrect result as a test
for our code.

The values of the elastic coefficients have been extracted from
simulations of 64 membranes starting at different initial
configurations. The initial configurations were generated by randomly
placing 500 lipids in two different layers with a vertical (along the
$z$ direction) separation $a$ (the size of the atoms) between them.
The initials configurations were ``thermalized'' over a period of
$2\times10^5$ MC time units, followed by a longer period of
$1.2\times10^6$ time units during which quantities of interest were
evaluated. The uncertainties in our final results correspond to one
standard deviation in the estimates of the averages. We first made the
simulations with non-fluctuating membranes, from which we extracted
the values of the Hamiltonian coefficients. Then, we removed the part
in our algorithm which is responsible for keeping the membrane flat.
The membranes were equilibrated again, and then the values of the
thermodynamic (free energy) coefficients were determined.

For the bare coefficients we find the following values for the surface
tension:
\begin{eqnarray}
\sigma_0&=&\left(0.8\pm 0.5\right)\ \frac{k_BT}{a^2}
\nonumber\\
\tilde{\sigma}_0&=&\left(-0.07\ \pm 0.01\right)\frac{k_BT}{a^2}.
\label{barestres}
\end{eqnarray}
The comparison of these results with each other, and with the values
of the elastic coefficients extracted from the fluctuation approach
[Eq.(\ref{fluctres})] reveals: (a) a disagreement between the two
surface tensions $\sigma_0$ and $\tilde{\sigma}_0$, which should be
attributed to the finite size of our membrane (see our discussion in
section \ref{sectension}); and (b) a disagreement between $\sigma_0$
and $\sigma$ which should be attributed to the entropic contribution
to the surface tension. The bending modulus $\kappa_0$ has been
obtained by dividing the system into $N_s$ stripes and extrapolating
the results for $\kappa_0$ to the limit $1/N_s\rightarrow 0$, as
explained earlier in this section (see also the Appendix). From the
extrapolation procedure, which is summarized in Fig.~\ref{extrap1}, we
find that
\begin{equation}
\kappa_0=\left(44\pm 10\right)\ k_BT.
\label{kappares}
\end{equation}
This result also serves as our estimate for $\kappa$ (see discussion
earlier in this section). The similarity of the above value of
$\kappa$ (which is, unfortunately, obtained with a somewhat large
numerical uncertainty) to the one quoted in Eq.(\ref{fluctres})
corroborates the argument presented in section \ref{routes} regarding
the equivalence of the two routes to membrane elasticity. Further
support to this argument is obtained from the agreement of our result
in Eq.(\ref{fluctres}) to $\sigma$, with the value of the surface
tension obtained from equilibrium approach [using expression
(\ref{stension1})]:
\begin{equation}
\sigma=\left(-0.3\pm 0.5\right)\ \frac{k_BT}{a^2}.
\end{equation}
Our result for $\tilde{\sigma}$ [expression (\ref{stension2})] is not
very much different
\begin{equation}
\tilde{\sigma}=\left(-0.41\pm 0.01\right)\ \frac{k_BT}{a^2}.
\end{equation}
These values are quite different from those given in
Eq.(\ref{barestres}), thus demonstrating the importance of the entropic
contribution to the surface tension.

Finally, we plot in Fig.~\ref{extrap2} our results for the
``apparent'' bending modulus $\kappa^*$ which we have obtained, using
expression (\ref{bending1}), from simulations of a fluctuating
membrane. These simulations serve as a test for our code. We find 
$\kappa^*=(-4\pm 8)\ k_BT$ which is consistent with the anticipated value
$\kappa^*=0$. 

\section{Summary and Discussion}
\label{summary}

Motivated by the lack of a well accepted theory to deal with the
statistical-mechanical behavior of curved interfaces, we have studied
the elastic properties of fluid bilayer membranes using analytical and
computational tools. Two distinct methods have been employed to
measure the surface tension $\sigma$, and the bending modulus
$\kappa$, of a model membrane. In the first (``fluctuation'') method
the elastic coefficients were extracted from the analysis of the
spectrum of thermal fluctuations of the membrane. The second
(``equilibrium'') method is based on the fact that $\sigma$ and
$\kappa$ describe the free energy variations due to area-changing and
curvature-forming deformations and, therefore, can be related to the
derivatives of the partition function with respect to the relevant
strain variables.  Using this kind of relation, we have derived formal
expressions for $\sigma$ and $\kappa$ in central force systems. Our
expressions associate the elastic coefficients to the interactions
between the molecules and the two-, three-, and four-particles
distribution functions. The most important feature of these
expressions is the fact that even though $\sigma$ and $\kappa$ (as
well as the saddle-splay modulus $\bar{\kappa}$) are related to
deformations of the membrane, they can be extracted from a single MC
run performed on the reference (unstrained) system.

One of the puzzles about curved interfaces elasticity is related to
the correspondence between the above two approaches for determining
their rigidity constants. We used linear response theory to prove that
the two methods must agree for the values of $\sigma$ and $\kappa$
provided that the system is deformed by the application of external
forces and not by altering other thermodynamic variables such as the
temperature or the chemical potential of surface molecules. Moreover,
our discussion clarifies that the coefficients in question, $\sigma$
and $\kappa$, are the {\em effective}\/ elastic coefficients which
appear in the Helfrich free energy (rather than the Helfrich
Hamiltonian) and which are influenced by the thermal undulations of
the membrane. Our computer simulations and the numerical values of the
elastic coefficients which we find, confirm the idea of equivalence
between the two routes to membrane elasticity.

Comparison of the computational efficiency of the two methods shows
that for our membrane model system the fluctuation method provides
more accurate estimates of the elastic coefficients than the
equilibrium method, and requires less CPU time. The major shortcomings
of the fluctuation approach is the fact that it can be utilized for
measurements of the effective coefficients only, and that it requires
the determination of the profile of the interface during the course of
the simulations. While this is easy with our ``water-free'' computer
model, this may not be so in other cases, for instance, for membranes
which tend to exchange molecules with the embedding solvent, or for
liquid-vapor interfaces near the critical point when the interface is
difficult to distinguish from the bulk phases. In these cases the
equilibrium method may be more attractive since the interactions in
the bulk phases do not contribute to the values of $\sigma$ and
$\kappa$ when calculated using our expressions for the elastic
coefficients. Moreover, with the same mechanical expressions for
$\sigma$ and $\kappa$, the bare (Hamiltonian) coefficients can be also
calculate. Our measurements demonstrate that close to the tensionless
state of the membrane, the entropic component of the surface tension
is quite significant. This has been also found recently in a
theoretical study of the surface tension of fluctuating surfaces
\cite{farago3}.

Finally, we would like to reemphasize that our expressions for the
elastic coefficients apply for central force systems only. Following
our derivation of these expression one should be able to generalized
them to more complicated cases including many-body interactions. A
more realistic model must also include electrostatic interactions
whose long-range nature pose a computational challenge.

{\em Acknowledgments:}\/ We thank Ram Seshadri for his comments on the
manuscript, and to Jeffrey Barteet for computational support in the
Materials Research Laboratory (MRL).  This work was supported by the
National Science Foundation under Award No.~DMR-0203755. The MRL at UC
Santa Barbara is supported by NSF No.~DMR-0080034.

\appendix

\section{Determination of {\protect $\kappa$} using the method of 
stripes}

The most common way to reduce finite size effects in computer
simulations is obtained by employing periodic boundary conditions,
namely by regarding the simulation cell as part of an infinite
periodic lattice of identical cells. When the range of the
interactions is less than $L_p/2$ (half the linear size of the cell)
than each particle $\alpha$ interacts only with the nearest periodic
image of any other particle $\beta$. This, in turn, is identified as
the pair $(\alpha,\beta)$. Each {\em pair}\/ has infinitely many
periodic images each of which is associated with a different
simulation cell; and with each simulation cell each pair is associated
exactly once. The set of all the different pairs associated with one
of the cells [say, the original (``primitive'') cell] is the one over
which the summation in expressions (\ref{stension1}) and
(\ref{stension2}) for the surface tension should be performed.

Things become more complicated when we try to evaluate the bending
modulus $\kappa$ using expression (\ref{bending1}). In this case,
coordinates associated with the location of the pair
($\bar{X}^{\alpha\beta}$ and $\bar{Y}^{\alpha\beta}$) appear in the
expression, and so it becomes necessary to decide which of the
periodic images of each pair is actually associated with primitive
simulation cell ($-L_p/2\leq x,y<+L_p/2$) over which the sum in
Eq.(\ref{bending1}) is performed. The intuitive candidate is the
periodic image with $-L_p/2\leq
\bar{X}^{\alpha\beta},\bar{Y}^{\alpha\beta}<+L_p/2$.  Making this
choice, however, is not the right convention. The correct way to
handle the summation in expression (\ref{bending1}) can be deduced
from our derivation of expression (\ref{bending3}) which is
independent of the location of the origin of axes. Following the
discussion that led from Eq.(\ref{bending1}) to Eq.(\ref{bending3}) it
becomes clear that: (a) each quartet
$((\alpha,\beta),(\gamma,\delta))$ must be reproduced exactly twice
from sums in Eq.(\ref{bending1}) [or once, if the quartets
$((\alpha,\beta),(\gamma,\delta))$ and
$((\gamma,\delta),(\alpha,\beta))$ are treated as different], and (b)
that the central coordinate of the quartet,
$(\bar{X}^{\alpha\beta,\gamma\delta},\bar{Y}^{\alpha\beta,\gamma\delta})$,
must lie inside the region of the primitive simulation cell. These
requirements can be perceived as is we have a periodic boundary
condition problem with the pairs playing the role of particles and
with $(\bar{X}^{\alpha\beta},\bar{Y}^{\alpha\beta})$ serving as the
coordinates of the pairs. What can also be learned from expression
(\ref{bending3}) is the fact that $\kappa$ is associated with {\em
  pair-pair}\/ correlations. Therefore, its accurate measurement is
difficult in systems whose linear $L_p<2\xi$, where $\xi$ is the
relevant correlation length. We proceed our discussion assuming that
our system is sufficient large and obeys the above criterion.

In order to calculate the third term in Eq.(\ref{bending1}) we divide
our system into an even number of stripes $N_s=2M$ ($M$-integer)
parallel to the $x$ axis, as shown in Fig.~\ref{stripesfig}. The
fourth term in Eq.(\ref{bending1}) is calculated in the same manner by
dividing the system into the same number of stripes parallel to the
$y$ axis. In addition to the primitive cell we also need to consider
the nearest periodic extensions of linear size $L_p/2$. These periodic
extensions, which are also shown in Fig.~\ref{stripesfig}, consist of
periodic images of the stripes. We label the stripes included in the
primitive cell with the numbers $M+1,\ldots,3M$, the stripes on the
right periodic extension with $1,\ldots,M$ (they are the periodic
images of stripes $2M+1,\ldots,3M$, and the stripes on the left
periodic extension (the images of stripes $M+1,\ldots,2M$) with
$3M+1,\ldots,4M$. For each pair we calculate the quantity
$p^{\alpha\beta}\equiv\phi'\left(R^{\alpha\beta}\right)R_x^{\alpha\beta}
R_z^{\alpha\beta}/R^{\alpha\beta}$. The location of the pair, which is
identified with the mid-coordinate
${\bar{X}}^{\alpha\beta}=\left(X^{\alpha}+X^{\beta}\right)/2$, defines
the stripe with which the pair should be associated. In
Fig.~\ref{stripesfig} each pair is depicted as a particle. The pair
labeled $a$, for instance, is located in the fifth stripe, whereas its
periodic image $a'$ is located in stripe number $13$. For each stripe
$i$ in the primitive cell we calculate the sum
\begin{equation}
\Sigma_i=\sum_{\rm pairs\ in\ stripe\ \#\ i} p^{\alpha\beta}
{\bar{X}}^{\alpha\beta}.
\end{equation}
The sum corresponding to stripe $j$, the image of stripe $i$, is given
by
\begin{equation}
\Sigma_j=\sum_{\rm pairs\ in\ stripe\ \#\ i} p^{\alpha\beta}
\left({\bar{X}}^{\alpha\beta}\pm L_p\right),
\end{equation} 
where the sign ($\pm$) in the above expression depends on whether the
image is situated to the right or the left of the primitive cell. The
product $\Sigma_p\Sigma_q$ gives the contribution to the third term in
Eq.(\ref{bending1}) of the quartets whose constituent pairs are
included, respectively, in stripes $p$ and $q$. These contributions
should be in accord with requirements (a) and (b), mentioned in the
previous paragraph, about the quartets and their locations. In some
cases these requirements are fulfilled by the image of the stripe
rather than the stripe itself. A few illustrative examples are given
in Fig.~\ref{stripesfig}: The contribution of the quartets $(a,b)$ and
$(b,c)$, for instance, is obtained from the products
$\Sigma_5\Sigma_8$ and $\Sigma_8\Sigma_{11}$, respectively. The
quartet $(a,c)$, on the other hand, should {\em not}\/ be introduced
into expression (\ref{bending1}) for $\kappa$ via the product
$\Sigma_5\Sigma_{11}$.  The distance from $a$ to the image $c'$ is
smaller than to $c$ and so the quartet should be identified as either
$(a,c')$ or as $(a',c)$.  The latter is the correct choice because the
center of the quartet $(a',c)$ satisfies
$-L_p/2\leq\bar{X}^{a',c}=\left(\bar{X}^{a'}+\bar{X}^c\right)/2<+L_p/2$,
while the center of the quartet $(a,c')$ falls outside the primitive
cell. The contribution to the expression for $\kappa$ of this pair is,
thus, obtained from the product $\Sigma_{11}\Sigma_{13}$.

The nice feature of the above examples is that the arguments we used
to reach our decisions about the correct way to handle the quartets
have {\em not}\/ been based on the {\em precise}\/ coordinates of the
pairs, but rather on the identity of the stripes and their locations
with respect to the center of the simulation cell. This means that the
products $\Sigma_p\Sigma_q$ reproduce the contribution of {\em all}\/
the quartets corresponding to the relevant stripes. Individual
decisions are necessary only for a small number of quartets,
associated with the following cases:
\begin{itemize}
\item{The first case is related with quartets in which the number of
    stripes separating the pairs is equal to $M$, as in the case of
    the pairs $b$ and $d$ in Fig.~\ref{stripesfig} which are located,
    respectively, inside the eighth and the twelfth stripes ($M=4$ in
    the above example). The separation between the pairs $b$ and $d$
    along the $x$ axis is very close to $L_p/2$, and it is impossible
    to know (without checking the coordinates of the pairs) whether
    the pair $d$ should be replaced by its periodic image $d'$ located
    in the fourth strips. In a homogeneous system about half of such
    pairs should be exchanged with their images, and so the best
    estimate for the contribution to expression (\ref{bending1}) for
    $\kappa$ arising from quartets including one pair inside the
    eighth stripe and the other inside the twelfth stripes is:
    $0.5\Sigma_8(\Sigma_4+\Sigma_{12})$.}
\item{Another case occurs when the stripes containing the two pairs
    are symmetric with respect to the center of the primitive cell
    and, in addition, the distance between them is larger than $M$. A
    typical example is the quartet $(a,d)$ in Fig.~\ref{stripesfig},
    in which $a$ is inside the fifth stripe and $d$ is in the twelfth
    strips. In this case it is obvious that $(a,d)$ has to be replaced
    by either $(a,d')$ or by $(a',d)$, but the two are equally
    probable.  Therefore, the contribution of such quartets is is best
    estimated by: $0.5(\Sigma_4\Sigma_5+\Sigma_{12}\Sigma_{13})$}
\end{itemize}
The above rules for correct summation over the different quartets can
be summarized by the following compact formula for the third term in
expression (\ref{bending1}):
\begin{equation}
\left\langle\sum_{p=1}^{4M}\sum_{q=1}^{4M}f_{p,q}\Sigma_p\Sigma_q
\right\rangle,
\label{kappaaprox1}
\end{equation}
where the function $f$ is given by
\begin{equation}
f_{p,q}=\left\{
\begin{array}{ll}
1 & {\rm for\ }|p-q|\leq M-1{\rm \ and\ } 2M+1<p+q<6M+1 \\
0.5 & {\rm for\ } |p-q|=M{\rm \ and\ } 2M+1<p+q<6M+1 \\
0.5 & {\rm for\ } |p-q|\leq M-1{\rm \ and\ }  p+q=2N+1 \\
0.5 & {\rm for\ } |p-q|\leq M-1{\rm \ and\ }  p+q=6N+1 \\
0 & {\rm otherwise}
\end{array} \right. .
\label{kappaaprox2}
\end{equation}
The value of $\kappa$ obtained using the above expressions
[Eqs.(\ref{kappaaprox1}) and (\ref{kappaaprox2})] are not accurate
since the contribution of some of the quartets is introduced in an
approximated way. However, the fraction of such quartets and the
resultant numerical error can be diminished by taking the limit
$N_s\rightarrow \infty$. In our simulations we have used a set of five
approximations with $N_s=4,6,8,12,24$.

{\em Another ``trick'' to speed up the calculation of $\kappa$:}\/ The
third and fourth terms in expression (\ref{bending1}) for $\kappa$
depend on the coordinates of the particles. Therefore, several values
for these quantities can be obtained from a single MC configuration by
generating replicas of the original simulation cell. These replicas
can be generated by shifting the position of the origin of axes, and
using the ``minimal image convention'' to define a replicated
primitive cell which is centered around the new origin. The
computational effort required for the calculation of expression
(\ref{bending1}) in the replicas is substantially smaller than that
required for the generation of new MC configuration. For one special
set of replicas the calculation can be done with (almost) no
additional effort at all: This set include the replicas generated when
the origin is shifted by constant intervals $\delta x=L_p/N_s$ in the
$x$ direction ($\delta y=L_p/N_s$ in the $y$ direction). Such shifts
are computationally favorable because they lead to cyclic permutations
of the stripes, but do not mix the pairs included in each one of them.


\newpage

\begin{figure}
  {\centering \hspace{1.5cm}\epsfig{file=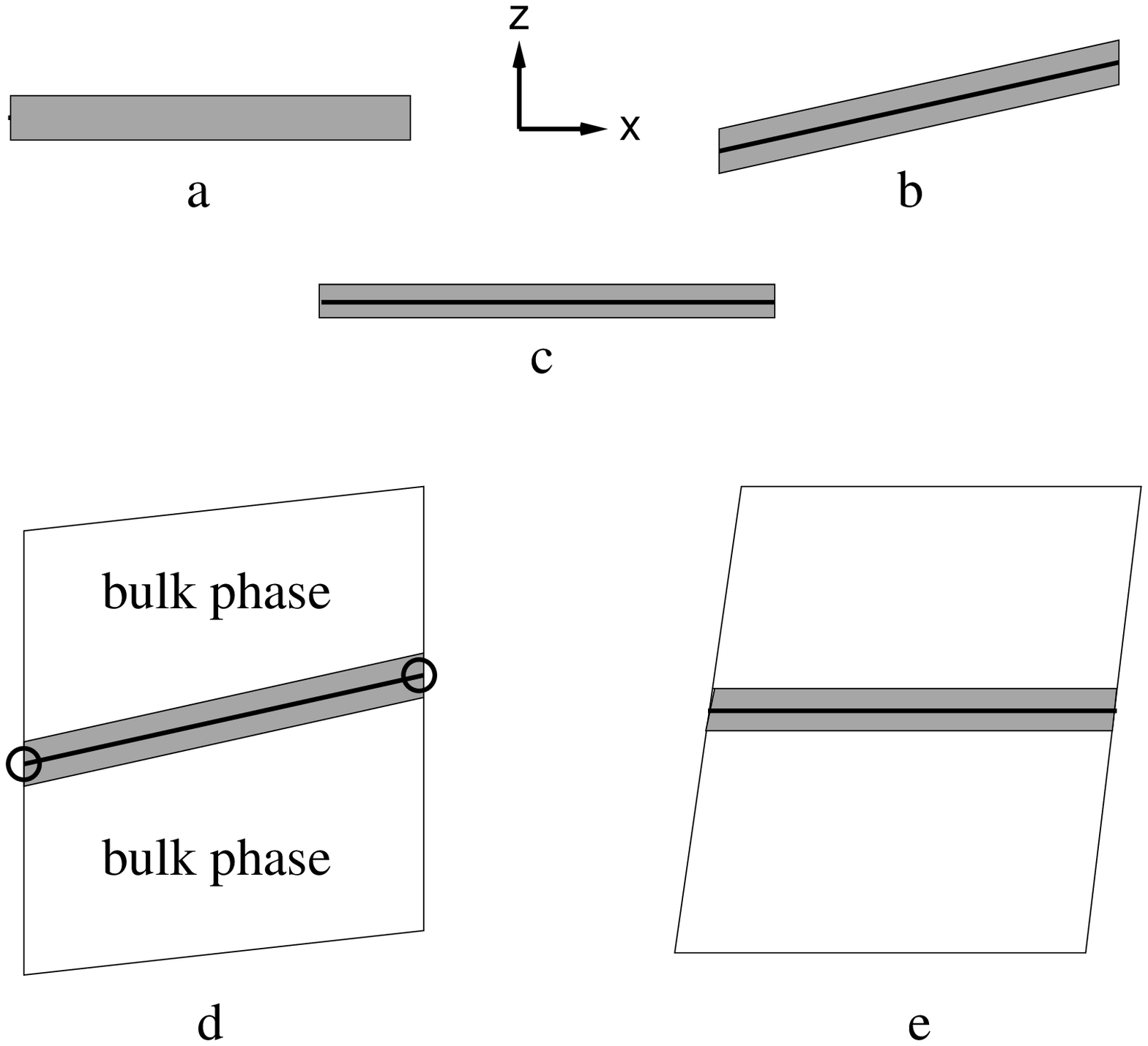,width=13cm}} 
\caption{A schematic picture of a bilayer membrane (gray) in the reference 
  state (a), and in two deformed states (b) and (c). The solid line
  represents the characteristic surface of the membrane, to which the
  Helfrich free energy is applied. The areas of the characteristic
  surfaces and the volumes of the membranes (represented by the
  gray-shaded area in the figure) in (b) and (c) are identical. The
  membrane depicted in (b) is shown in (d) together with the
  containing cell and the embedding solvent. The end points marked by
  the open circles belong to the perimeter {\protect ${\cal P}$} of
  the characteristic surface.  Another deformation of the container,
  which do not change the total area of the characteristic surface, is
  shown in (e).}
\label{membrane1}
\end{figure}

\begin{figure}
  {\centering \hspace{1.5cm}\epsfig{file=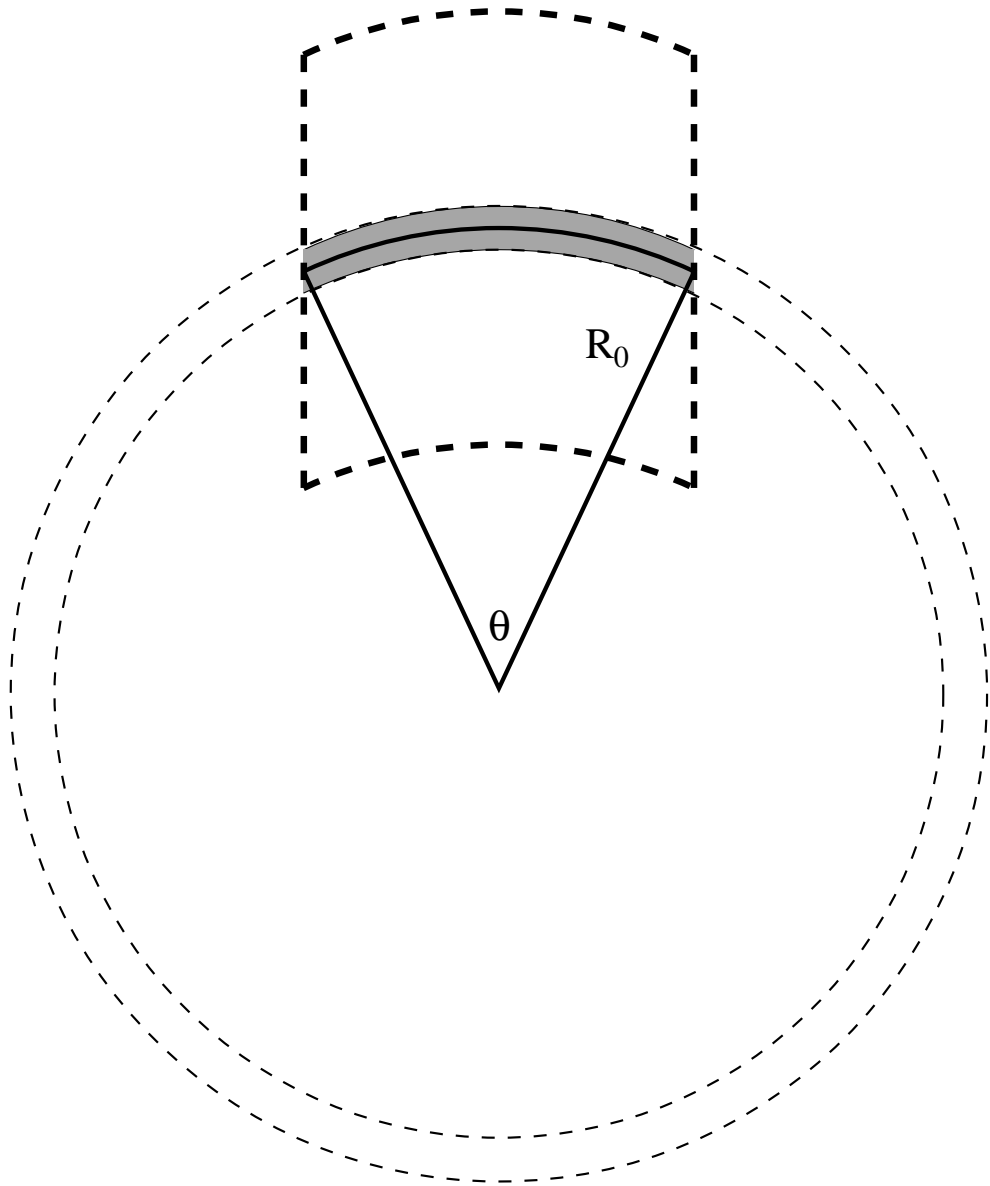,width=13cm}} 
\caption{A cylindrical bilayer membrane (gray) with radius of curvature 
  {\protect $R_0$} and apex angle {\protect $\theta$}.  The solid line
  in the middle of the membrane represents the characteristic surface.
  The cylindrical shape of the membrane is obtained via a deformation
  of the containing cell, depicted by the bold dashed line in the
  figure. The membrane may be thought of as part of a cylindrical
  vesicle (depicted by the thin dashed line) of a similar radius of
  curvature.}
\label{membrane2}
\end{figure}

\begin{figure}
  {\centering \hspace{1.5cm}\epsfig{file=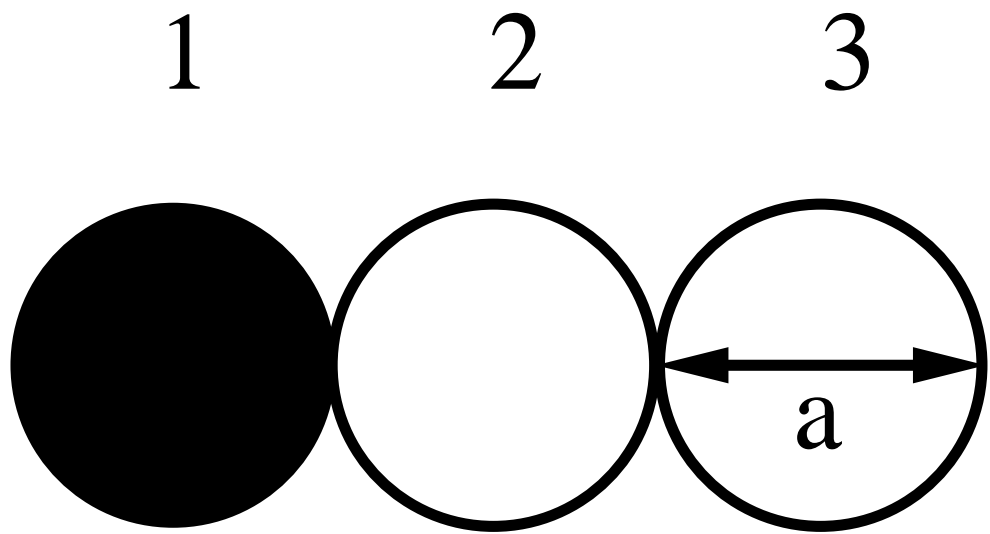,width=13cm}} 
\caption{A schematic picture of a lipid molecule in our model system - a
  trimer consisting of three spherical atoms of diameter {\protect
    $a$}.  The atom labeled 1 (solid circle) represents the
  hydrophilic head of the lipid, while the atoms labeled 2 and 3 (open
  circles) represent the hydrophobic tail.}
\label{lipid}
\end{figure}

\begin{figure}
\vspace{4cm} 
{\centering\hspace{1.5cm}\epsfig{file=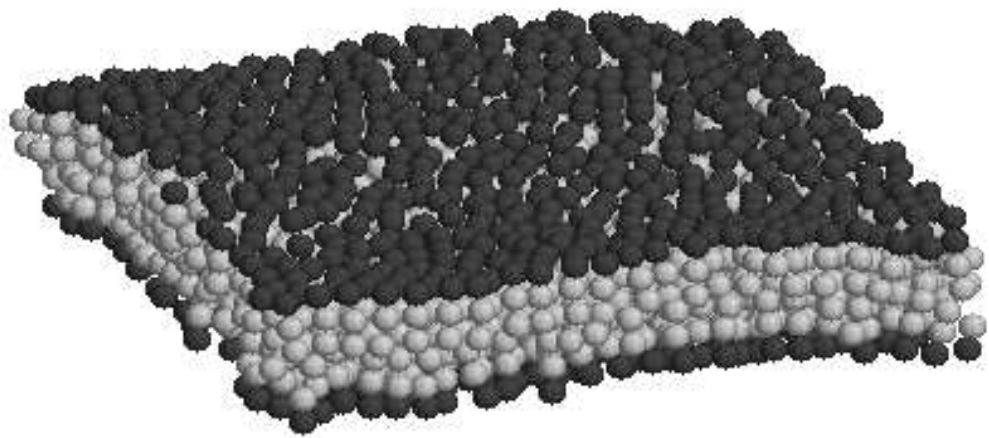,width=13cm}}
\caption{Equilibrium configuration of a fluid membrane consisting of 1000 
  molecules (500 molecules in each monolayer).}
\label{membranefig} 
\end{figure} 

\begin{figure}
{\centering \hspace{1.5cm}\epsfig{file=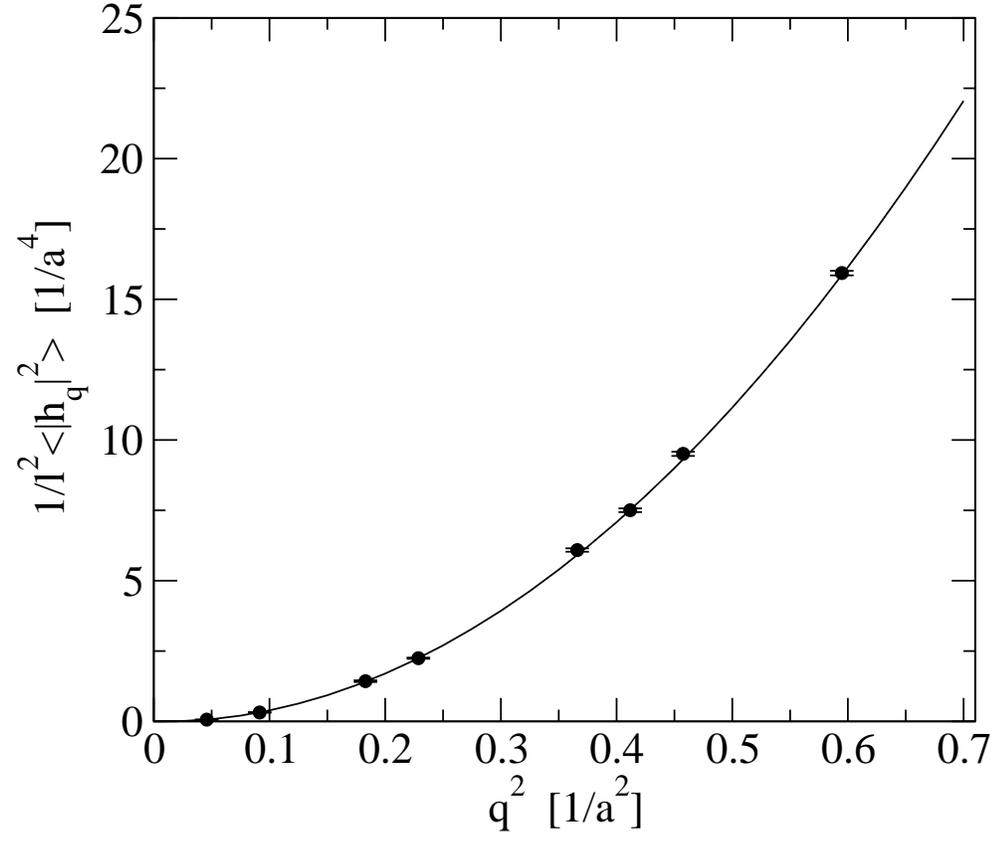,width=13cm}}
\caption{The inverse of the spectral intensity for undulatory modes 
  {\protect $1/l^2\langle|h_q|^2\rangle$} as a function of the square
  wave number {\protect $q^2$}. The circles mark numerical results,
  while the solid line depicts {\protect Eq.(\ref{fluctexpress2})} with
    the values of {\protect $\sigma$ and $\kappa$} given by {\protect
      Eq.(\ref{fluctres})}}.
\label{spectrum}
\end{figure}

\begin{figure}
{\centering \hspace{1.5cm}\epsfig{file=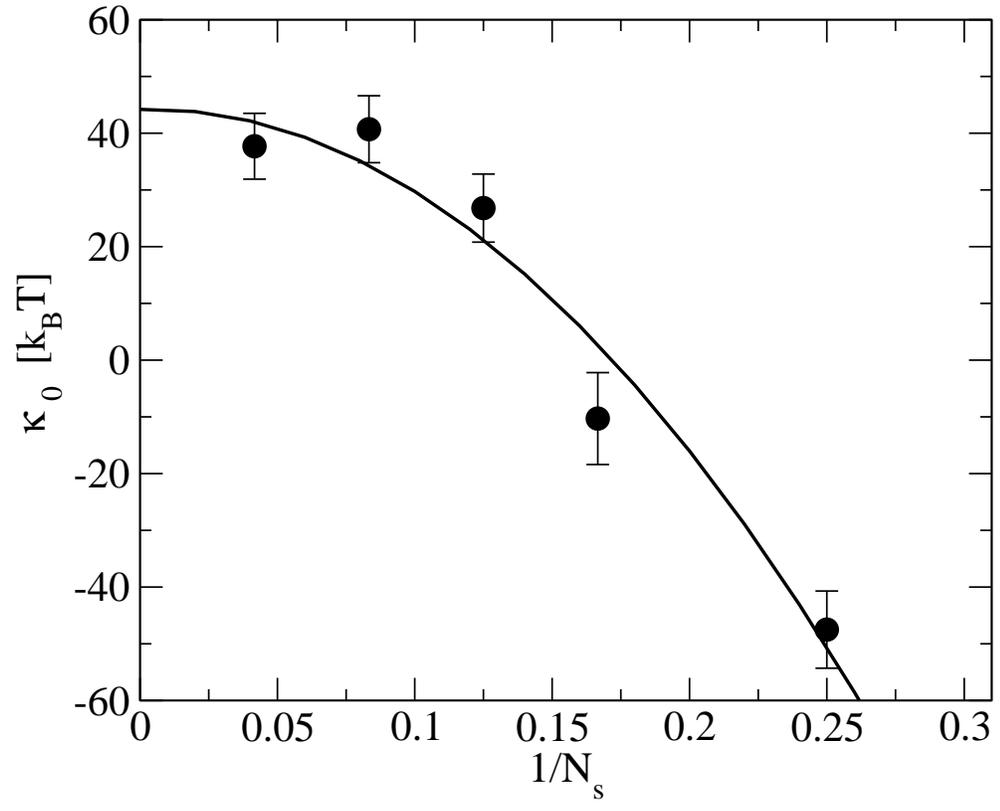,width=13cm}}
\caption{The bending modulus $\kappa_0$ as a function of the inverse of 
  number of stripes dividing the simulation cell, $1/N_s$. The curve
  depicts the weighted least square fit of a second order polynomial
  in $1/N_s$ to the data.}.
\label{extrap1}
\end{figure}

\begin{figure}
{\centering \hspace{1.5cm}\epsfig{file=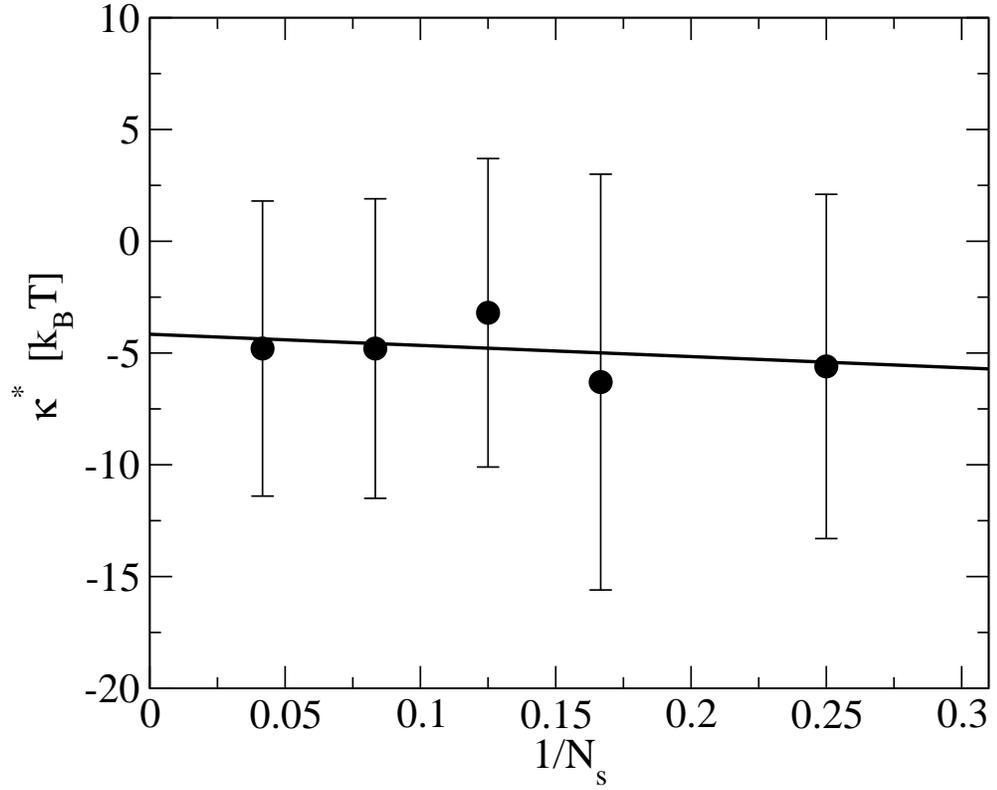,width=13cm}}
\caption{The ``apparent'' bending modulus $\kappa^*$ as a function of the 
  inverse of number of stripes dividing the simulation cell, $1/N_s$.
  The curve depicts the weighted least square fit of a first order
  (linear) polynomial in $1/N_s$ to the data.}.
\label{extrap2}
\end{figure}

\begin{figure}
{\centering \hspace{1.5cm}\epsfig{file=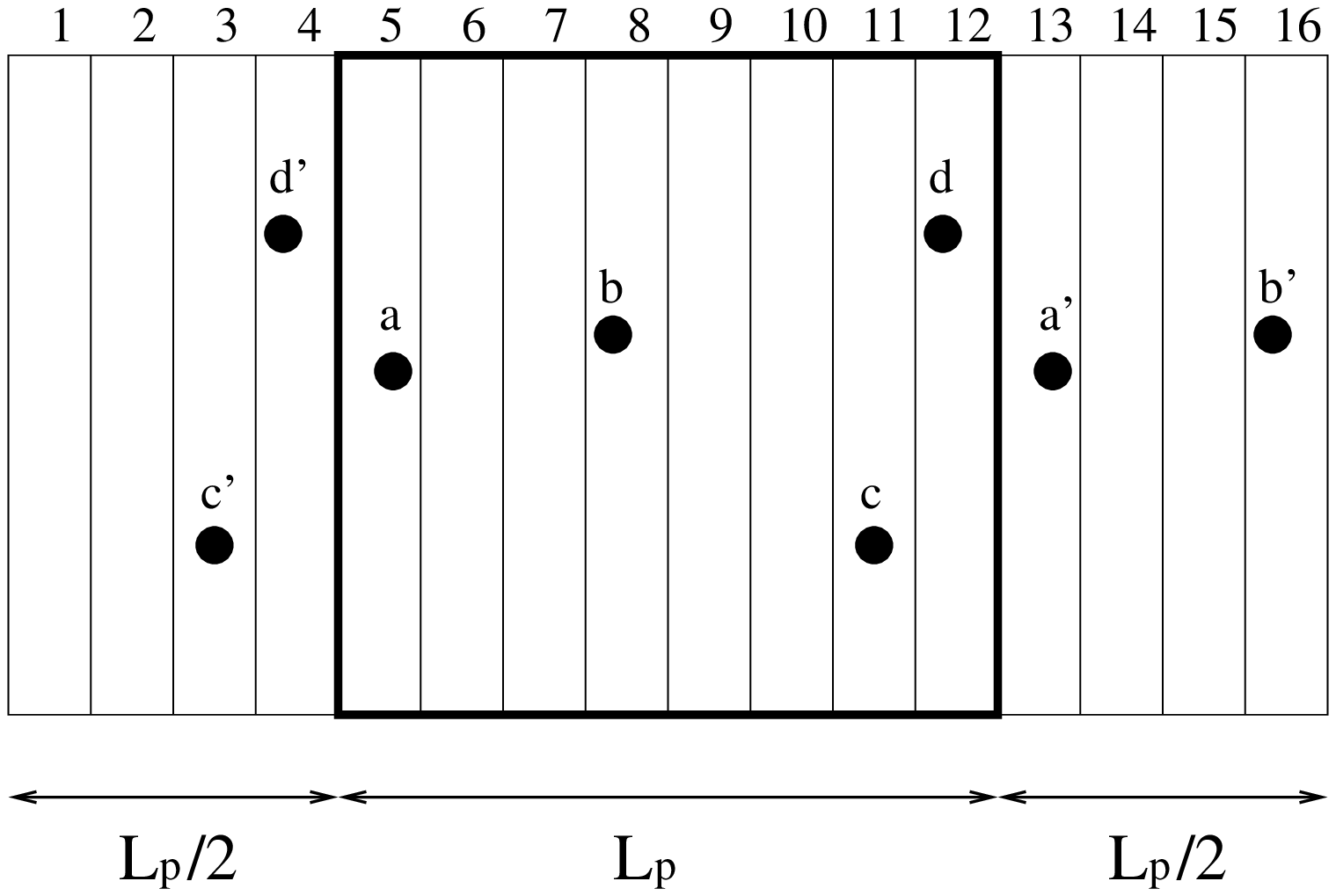,width=13cm}}
\caption{A schematic picture of a system of linear size $L_p$ consisting of 
  four pairs (a,b,c,d) and their periodic images (a'.b',c',d'). The
  bold frame marks the boundaries of the primitive simulation cell
  which is divided into $N_s=8$ stripes labeled from 5 to 12. The
  images of the stripes which belong to the nearest periodic
  extensions of the primitive cell are labeled 1-4 and 13-16}
\label{stripesfig}
\end{figure}

\end{widetext}

\begin{thebibliography}{99}
\bibitem{remarkforces} We would like to remind the reader that, in
  general, the area of the reference state does not coincide with the
  Schulman area which is the energy ground state of the system. The
  surface pressure which must be applied to fix the area of the
  surface at the reference state is nothing but the negative of the
  surface tension.
\bibitem{remarktrans} We do not argue that the coordinates of atoms
  deform affinely, but rather use transformation {\protect
    (\ref{lintransform})} as a mapping between the configuration phase
  spaces of the reference and the deformed systems.
\bibitem{remarkade} A classical example is the discussion on
  elasticity of thin plates in Ref.~{\protect \cite{landau_lifshits}}
  (which has inspired many other works), where the area-difference
  rather than the bending energy is calculated.
\end{thebibliography}
\end{document}